# Metalens stereoscopic optical palpation for imaging cancer mechanics

Haoyi Yu[1], Rhys Jones[2,3], Chi Li[1], Bo He[4], Renate Zilkens[2,5], Louise N. Winteringham[6], Laura Gale[7,8], Mireille Hardie[7,8], Fiona Bellamy[9], Saud Hamza[9], Christobel M. Saunders[5,10,11], Liver Cancer Collaborative[§], Stefan A. Maier[1,12], Brendan F. Kennedy[2,3,13], Haoran Ren[1,*], and Qi Fang[2,3,*]

[1]*School of Physics and Astronomy, Faculty of Science, Monash University, Melbourne, Victoria 3800, Australia.*

[2]*BRITElab, Harry Perkins Institute of Medical Research, QEII Medical Centre, Nedlands and Centre for Medical Research, The University of Western Australia, Crawley, WA 6009, Australia*

[3]*Department of Electrical, Electronic & Computer Engineering, School of Engineering, The University of Western Australia, Crawley, WA, 6009, Australia*

[4]*Cancer Microenvironment Laboratory, Harry Perkins Institute of Medical Research, Centre for Medical Research, The University of Western Australia, Perth, WA, Australia*

[5]*Medical School, The University of Western Australia, Crawley, WA 6009, Australia*

[6]*Harry Perkins Institute of Medical Research, QEII Medical Centre, Nedlands and Centre for Medical Research, The University of Western Australia, Crawley, WA 6009, Australia*

[7]*School of Pathology and Laboratory Medicine, The University of Western Australia, Crawley, WA 6009, Australia*

[8]*PathWest, Fiona Stanley Hospital, 11 Robin Warren Drive, Murdoch, WA, 6150, Australia*

[9]*Breast Centre, Fiona Stanley Hospital, 11 Robin Warren Drive, Murdoch, WA, 6150, Australia*

[10]*Department of Surgery, Melbourne Medical School, The University of Melbourne, Parkville, Victoria 3010, Australia*

[11]*The Royal Melbourne Hospital, 300 Grattan Street, Parkville, Victoria, 3050, Australia*

[12]*Department of Physics, Imperial College London, London SW7 2AZ, U.K.*

[13]*Institute of Physics, Faculty of Physics, Astronomy and Informatics, Nicolaus Copernicus University in Toruń, ul. Grudziądzka 5, 87-100 Toruń, Poland*

[§] *A list of authors and their affiliations appear at the end of the paper*

[*]Corresponding author: haoran.ren@monash.edu

qi.fang@uwa.edu.au

## Abstract

Accurate intraoperative margin assessment is critical for complete resection of solid tumours. However, surgeons routinely rely on manual palpation, which is highly subjective and fundamentally limited by poor spatial resolution, and pre-operative imaging lacks real-time feedback. While emerging optical elastography techniques offer quantitative mechanical contrast, their reliance on bulky optical components necessitates cumbersome alignment and hinders integration into miniaturised surgical instruments. Here, we report metalens stereoscopic optical palpation (MSOP), a nanophotonic imaging platform leveraging a binocular metalens integrated with a single CMOS sensor to achieve high-fidelity, self-aligned elastography. By replacing complex bulk optics with a compact stereoscopic architecture, MSOP enables robust, high-contrast mapping of tissue surface stress. Following validation in heterogeneous silicone phantoms, we demonstrate MSOP's translational potential by characterising the mechanical signatures of malignancies in fresh mouse pancreatic cancer models, as well as excised human breast and liver specimens. By providing reliable, label-free mechanical contrast across diverse oncological landscapes, MSOP offers a compact tool for precise intraoperative margin delineation to reduce re-excision rates and improve surgical outcomes.

Cancer remains a leading cause of global mortality, responsible for almost one in six deaths worldwide (1). Surgical resection serves as the primary curative-intent treatment for early-stage solid tumours (2, 3). During these procedures, surgeons frequently rely on manual palpation to identify neoplastic margins, a technique based on the distinct mechanical contrast between stiff malignant lesions and more compliant benign tissue (4-8). Despite its ubiquity, manual palpation is fundamentally limited by its subjective nature and poor spatial resolution, both of which are constrained by the human tactile threshold (5, 9, 10). Consequently, sub-centimetre or deep-seated lesions often remain undetectable due to the finite sensitivity of human mechanoreceptors. This tactile blind spot often results in positive surgical margins, leading to incomplete resections and increased rates of recurrence (6, 9, 11-13).

To overcome the limitations of manual palpation, optical elastography has emerged as a powerful paradigm for digitising tactile feedback and mapping tissue mechanical contrast with high spatial resolution (14-16). While non-optical tactile sensors offer direct mechanical quantification, they are frequently hampered by wiring complexities and limited spatial resolution that restrict their broader clinical adoption (17-20). Although optical elastography circumvents these constraints, its clinical translation is fundamentally bottlenecked by system complexity and a reliance on conventional refractive optics. For instance, optical coherence elastography provides quantitative mapping of mechanical properties but demands expensive, intricate interferometric setups (21). Alternatively, cost-effective variants such as coherence function-encoded (22) and camera-based optical palpation (23) rely on opaque, compliant stress-sensing layers that obscure the underlying anatomy, hindering the precise co-registration of the generated stress maps with specific tissue features. Recent development of stereoscopic optical palpation (SOP) systems circumvents this opacity by utilising transparent elastomers with embedded phosphorescent microparticles for sequential visual imaging and optical palpation. However, these systems intrinsically rely on dual-camera configurations (24, 25), resulting in a bulky form factor that complicates seamless integration into spatially constrained minimally invasive instruments. Moreover, these dual-sensor SOP implementations necessitate rigorous extrinsic calibration and introduce inter-sensor synchronisation errors that can compromise diagnostic reliability. Consequently, overcoming the physical constraints of conventional bulk optics remains a critical hurdle for the broader clinical translation of optical elastography.

Metasurfaces (26, 27), which are ultrathin planar arrays of sub-wavelength meta-atoms, have emerged as a transformative photonic platform in wavefront engineering, providing a pathway to overcome these geometric optical constraints. By locally tailoring the phase, amplitude, and polarisation of incident photons, these quasi-two-dimensional surfaces can consolidate the functionality of bulky, multi-element optical trains into a single, chip-scale footprint (28, 29). Notably, ultrathin metalenses have recently been engineered for diffraction-limited (29), achromatic (30, 31), metafibre (32), and stereoscopic imaging (33-36), successfully bypassing the inherent physical limitations of traditional refractive lenses. In the biomedical domain, metasurfaces are increasingly leveraged to miniaturise complex modalities, ranging from endoscopic metalenses to point-of-care holographic sensors (37-39).

Here, we report the development of metalens stereoscopic optical palpation (MSOP), capable of high-contrast elastography and delineation of cancerous tissue without the geometric, electronic, and optical overhead of dual-camera systems. By integrating binocular metalens directly with a single complementary metal–oxide–semiconductor (CMOS) sensor, we demonstrate a compact system that captures stereoscopic views simultaneously, ensuring inherent temporal synchronisation and eliminating the need for complex calibration. This highly integrated, single-sensor approach reduces the overall instrument volume by more than a factor of ten, thereby enabling potential integration into minimally invasive surgical tools. After validating the system’s ability to reveal mechanical contrast in heterogeneous silicone phantoms, we demonstrated its clinical potential by mapping the diverse mechanical profiles of mouse pancreatic cancer models and human breast and liver cancer specimens. Our results reveal that MSOP provides robust mechanical contrast, clearly delineating malignant lesions from surrounding benign tissue to facilitate accurate intraoperative margin assessment.

## Results

### MSOP system setup and working principle

To enable robust and label-free mechanical mapping of biological tissues, we developed a benchtop MSOP system. The system setup and its fundamental working principle are detailed in Fig. 1. As illustrated in Fig. 1a, the MSOP imaging head is housed within a custom-engineered, multi-component 3D-printed chassis with an overall size of 6×6×6 $cm^3$. The assembly comprises an outer shell with an integrated basal transparent glass window, a middle housing containing a binocular metalens with adjustable axial positioning, and an inner compartment securing the CMOS sensor. To ensure precise focus control, the CMOS sensor is mounted via a screw-and-spring mechanism, allowing for fine-tuned adjustment of the sensor-to-metalens distance. The entire imaging head is mounted on a manual $z$-axis stage to accommodate varying sample geometries, while the sample itself is positioned on a high-precision 5-axis stage.

The stress sensing mechanism relies on a 5.2-mm compliant silicone layer embedded with phosphorescent microparticles at its bottom surface, which is interfaced between the glass window and the sample. During operation, the $z$-stage translates the imaging head to compress the sample-layer stack, typically reaching a bulk strain of 10–30%. The binocular metalens, comprising sub-wavelength nanopillar arrays (see Fig. 1a inset), projects two distinct views of the compressed layer onto a single CMOS sensor. This stereoscopic configuration directs the left and right imaging channels onto corresponding halves of the sensor plane. Crucially, the mechanical heterogeneity of the underlying tissue dictates the local deformation of the compliant layer; regions overlying stiff inclusions, such as tumours, induce greater local strain in the layer compared to those over compliant benign tissue.

The computational reconstruction pipeline begins with the sequential acquisition of dual-mode images, each captured in a single-shot exposure (Fig. 1b). Using mouse pancreatic tissue as a representative model, we first captured stereoscopic photographs under white-light illumination to visualise tissue morphology, followed by a corresponding UV-illuminated capture to resolve the phosphorescent microparticle distribution. To mitigate optical noise arising from UV scattering at the glass-tissue interface, the raw UV-illuminated images were processed through a digital green-colour filter. This step effectively isolates the discrete particle emissions, significantly enhancing the signal-to-noise ratio for subsequent digital image correlation (DIC). We implemented an open-source Matlab DIC framework (Ncorr, v1.2) to compute the disparity map. The filtered UV-illuminated stereoscopic pair was split into a reference (left) and current (right) image. A circular subset window (500 μm in diameter) was defined in the reference image and computationally tracked across the current image. The correlation algorithm converged when the cost function reached a threshold of $10^{-4}$ pixels. At this point, the horizontal component of the 2D displacement vector (the $x$-axial pixel shift parallel to the metalens baseline) was recorded as the local disparity, as the system's integrated design ensures that vertical component remains negligible. By iterating this cross-correlation process across the entire field of view (FOV) of 9 mm in diameter, we generated a continuous disparity map representing the three-dimensional (3-D) deformation of the compliant layer.

The final stage of the MSOP pipeline involves the conversion of optical disparity into a quantitative map of tissue surface stress (Fig. 1c). Within the 9-mm FOV, the disparity map exhibits distinct spatial variations, with elevated localised disparity corresponding to stiffer underlying tumour regions. To quantify this relationship, we performed a calibration by compressing the compliant layer against a rigid, flat substrate at increments of 500 μm. The resulting disparity-strain response was modeled using a second-degree polynomial fit (Fig. 1c, plot with red curve). By applying this calibration to the measured disparity map, we derived an intermediate strain map. Finally, by integrating the pre-characterised uniaxial stress-strain relationship of the silicone layer (Fig. 1c, plot with blue curve), we transformed the strain distribution into a quantitative stress map. Under the assumption of uniaxial stress distribution and continuity across the layer-sample interface, this reconstructed stress map represents the mechanical contrast of the tissue surface, providing an objective diagnostic signature of the tumour's physical boundaries. This entire MSOP computational pipeline, from raw stereoscopic image acquisition to stress map reconstruction, is highly efficient, requiring approximately one minute using a standard personal laptop.

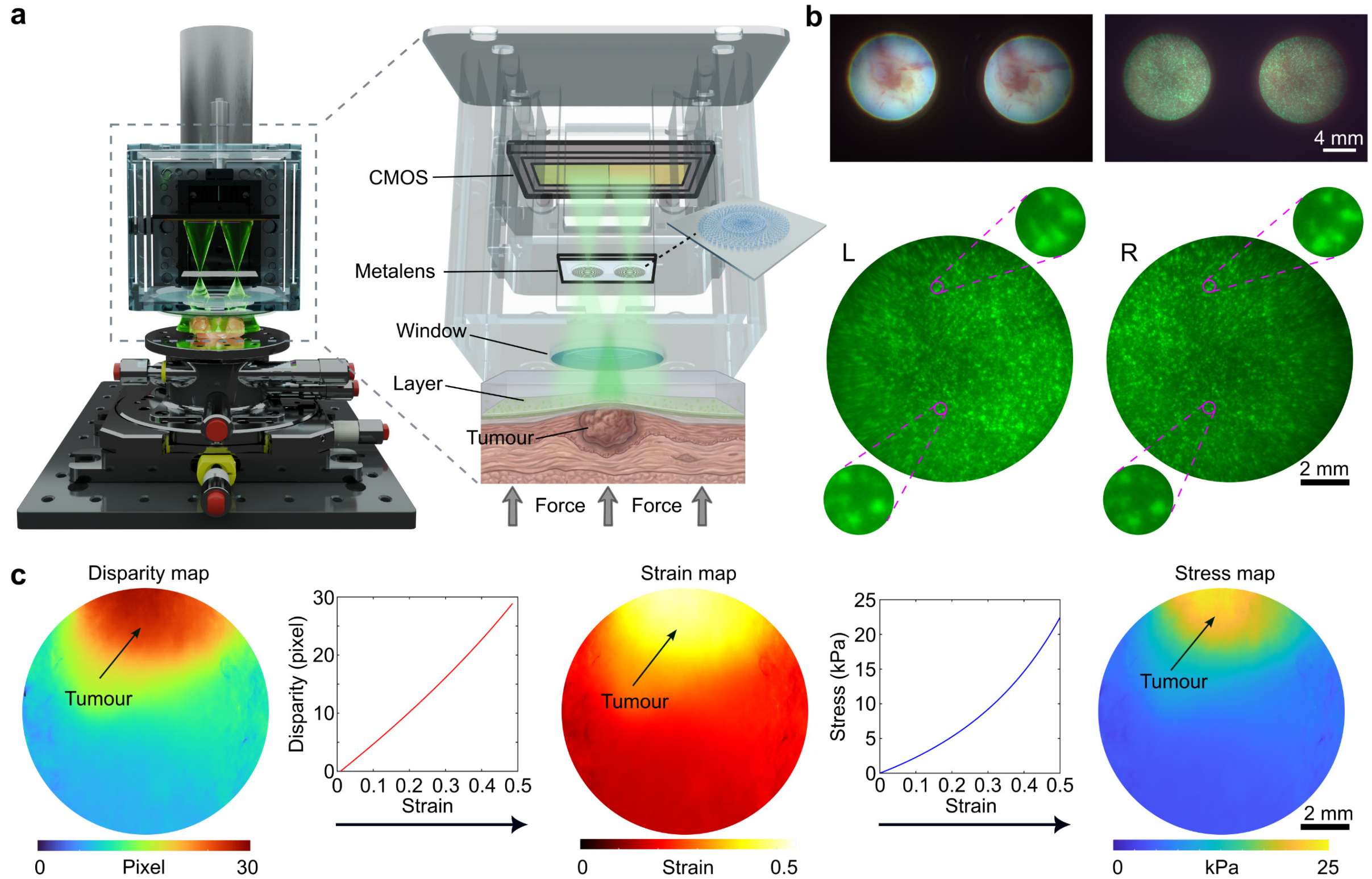

**Figure 1**. **Experimental setup, working principle, and image processing pipeline. a** Schematic illustration of the benchtop MSOP experimental configuration. The zoomed-in perspective depicts the optical path where light emitted from the phosphorescent stress-sensing layer is captured by the binocular metalens and projected as a stereoscopic pair onto a single CMOS sensor. The inset provides a representative view of the sub-wavelength nanopillar arrays comprising the metalens. **b** Image acquisition and DIC workflow. The upper row shows raw stereoscopic images of mouse pancreatic tissue under dual-mode illumination: white light for morphological visualisation and UV light for phosphorescent particle excitation. The lower row displays spectrally filtered (green channel) images used to isolate particle features for DIC. Magenta circles indicate the representative 500-μm subset windows used for feature tracking and disparity calculation. **c** Quantitative mechanical mapping pipeline. The process begins with the generation of a disparity map from the UV-illuminated stereoscopic pair. This optical data is converted into an intermediate strain map using a pre-calibrated disparity–strain relationship (first inset plot, red curve). Finally, the strain map is transformed into a quantitative stress map by applying the uniaxial stress-strain characterisation of the compliant stress-sensing layer (second inset plot, blue curve), revealing the mechanical contrast of the sample.

## Design and characterisation of the binocular metalens

The binocular imaging module was designed by first prescribing the phase response required for each individual metalens to transform an incident plane wave into a spherical wave converging to the designed focal plane. For a metalens with focal length $f = 4\,mm$, the target hyperbolic phase profile is given by $\phi(r,\lambda) = -\frac{2\pi}{\lambda_0}\left(\sqrt{r^2+f^2}-f\right)$, where $r$ is the radial coordinate and $\lambda_0$ is the design wavelength. The phase profile was wrapped into the range of $0$ to $2\pi$ and implemented using cylindrical $Si_3N_4$ nanopillars with spatially varying diameters. The nanopillar diameters were selected from a simulated phase library to provide full $2\pi$ phase coverage while maintaining high transmission at the design wavelength of 550 nm.

Figure 2a illustrates the stereoscopic imaging module, which comprises a pair of $Si_3N_4$ metalenses integrated with a single CMOS sensor. The two metalenses feature a lateral baseline separation of 3.4 mm, enabling two distinct parallax viewpoints to be captured simultaneously in a single exposure for disparity-based depth sensing. Each metalens forms an independent image channel on the sensor, yielding a calibrated object-plane FOV of 10.89 mm in diameter per channel. Notably, this raw optical FOV is intentionally larger than the final 9-mm effective FOV of the MSOP system; this design buffer allows the peripheral boundaries of the reconstructed stress maps to be cropped, thereby eliminating edge artefacts caused by peripheral optical distortion. The binocular metalens is mounted within a compact 3D-printed housing, which accommodates the glass-substrate metalens, a thin-film green-colour filter, and the CMOS sensor.

Figure 2b depicts the fabricated $Si_3N_4$ metalens with a diameter of 1 mm. The design employs cylindrical nanopillars with a fixed height of h = 750 nm and a lattice period of p = 350 nm. The pillar diameters are selected to achieve full $0-2\pi$ phase coverage while sustaining simulated transmission above 90% at 550 nm (Fig. S1). Figure 2c shows the characterisation results obtained using a custom optical setup at 550 nm (Fig. S2). A supercontinuum source (NKT Photonics, SuperK FIANIUM) was spectrally filtered and directed onto the metalens from the substrate side. Axial intensity distribution was measured by translating the sample along the optical axis with a precision motorised stage and recording focal-plane profiles sequentially. The measured focal distance of 4.1 mm is in close agreement with the designed 4 mm focal length, with the minor deviation likely due to imperfect collimation of the incident beam. The measured transverse full width at half maximum (FWHM) of 2.35 μm closely matches the simulated diffraction-limited value of 2.24 μm. The focusing efficiency is 70.24%, calculated by integrating the focal-spot power at the focal plane and normalising to the incident power within the metalens aperture.

The imaging performance of the fabricated metalens was experimentally assessed using the USAF resolution test chart. Groups 4, 5, and 6 are resolved within the FOV, while features as fine as 2.76 μm linewidth (Group 7, Element 4) are distinguishable (Fig. 2d), consistent with the lateral resolution predicted from the focal-spot characterisation in Fig. 2c.

The contrast-transfer characteristics were further evaluated from the USAF-1951 target images shown in Fig. 2d, using a custom optical setup (Fig. S3). For each resolved element, the contrast transfer function (CTF) was calculated from the line-pair modulation as $\mathrm{CTF} = \frac{I_{max,i}-I_{min,i}}{I_{max,i}+I_{min,i}}$ where $I_{max}$ and $I_{min}$ were extracted from the bright and dark bars within the selected region of interest (ROI). The experimental CTF was then compared with a numerical CTF obtained from simulated USAF images generated using an angular-spectrum-based scalar diffraction model (40) (Fig. 2e). In the simulation, the imaging point-spread function (PSF) was determined based on the metalens imaging configuration, where both the object and image planes were positioned at 2f relative to the metalens plane. The USAF object pattern was then convolved with this PSF to generate the simulated image, which was subsequently sampled according to the pixel number and resolution of the CMOS sensor. Both the experimental and simulated CTF curves are in good overall agreement and exhibit a decrease with increasing spatial frequency, as expected from the combined factors of finite numerical aperture and pixelated sampling of the CMOS sensor. A local drop in the measured CTF is observed in the spatial-frequency range of 40–50 lp mm$^{-1}$, whereas the simulated CTF decreases more smoothly. This discrepancy is likely attributable to fabrication-induced phase errors. Overall, the good agreement confirms that the CMOS-integrated metalens preserves high-spatial-frequency information and maintains high imaging fidelity at the target wavelength. Further details are provided in Supplementary Note 3 and Fig. S4.

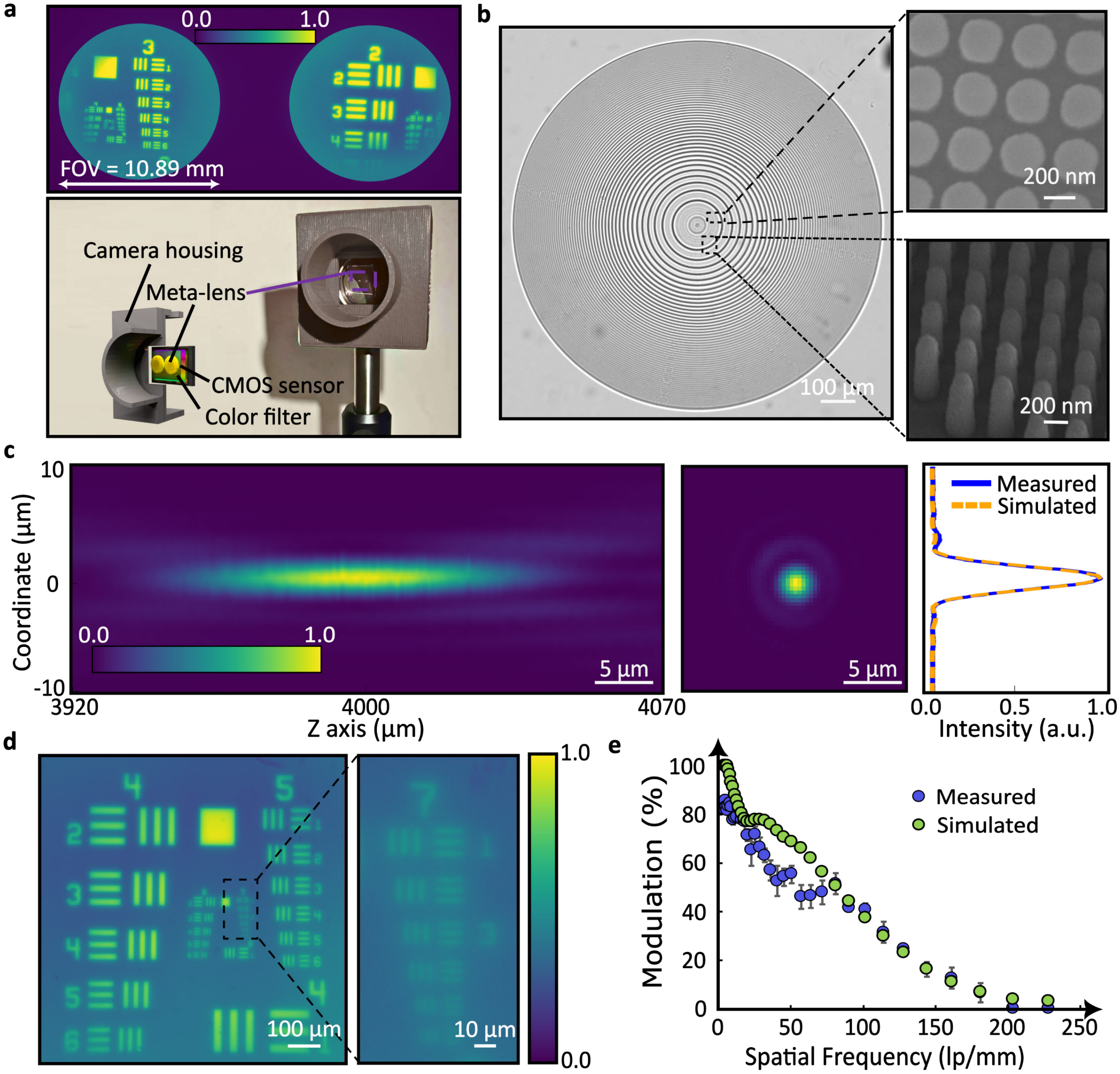

**Figure 2. Design, fabrication, and characterisation of a CMOS sensor-integrated binocular metalens. a** Photograph of the binocular metalens integrated with a CMOS sensor via a 3D-printed plastic housing. The lower schematic illustrates the internal layout of the imaging module. The circularly cropped stereoscopic USAF-1951 images shown above were captured through the two metalenses, with each circular image corresponding to a calibrated object-plane field of view (FOV) of 10.89 mm in diameter. The colour bar denotes normalised intensity. **b** Optical micrograph of a single metalens within the binocular configuration, featuring a 1-mm diameter $Si_3N_4$ platform. SEM images (top and side views)

reveal cylindrical nanopillars with diameter variations. **c** Experimentally measured focal-spot intensity distributions in both longitudinal and transverse planes at a wavelength of 550 nm. The left panel shows the longitudinal beam around the focal region. The middle panel shows the transverse point-spread function (PSF) measured at the focal plane. The right panel shows normalised line profiles across the focal spot, giving an experimental full width at half maximum of 2.35 μm, in close agreement with the simulated diffraction-limited value of 2.24 μm. The three panels share the same vertical spatial coordinate. **d** Imaging performance of the metalens measured using a USAF-1951 resolution target (Group 4 – 7) at 550 nm. The magnified region highlights the high-spatial-frequency features in Group 7. The colour bar denotes normalised intensity. **e** Contrast transfer function (CTF) measured from USAF-1951 images and compared with numerical simulation. Experimental data points represent the mean CTF extracted from five independent regions of interest within each USAF element, and error bars denote one standard deviation across these regions of interest. The simulated CTF was extracted from numerically generated USAF images obtained by convolving the USAF object pattern with the imaging PSF calculated for the corresponding object–image geometry, followed by pixel sampling by the used CMOS sensor.

## System validation and characterisation via silicone phantoms

To quantify the mechanical sensitivity and spatial precision of MSOP, we validated the system using heterogeneous silicone phantoms containing an embedded macroscale cylindrical inclusion (2 mm diameter) of higher stiffness than the surrounding base matrix (Fig. 3). The details of the silicone phantoms are described in Methods. We first established the operational pipeline by analysing a representative evaluation sequence under a 30% axial preload strain (Figs. 3a–f). Each complete evaluation sequence comprises a consistent suite of data products: white-light morphological photography (Fig. 3a), filtered UV-illuminated imaging to isolate the phosphorescent emissions of the stress-sensing layer (Fig. 3b), and the resulting reconstructed stress map (Fig. 3c). To evaluate the spatial fidelity of the reconstructed stress distributions, we extracted transverse stress profiles along the horizontal and vertical axes (white dash-dotted lines in Fig. 3c), as shown in Figs. 3d and 3e, respectively. An automated edge-localisation algorithm was implemented by fitting a first-order Gaussian derivative to the stress gradient, whereby the transverse peak-to-valley distance of this derivative provides an objective, operator-independent measurement of the effective inclusion width (yellow dashed vertical lines in Figs. 3d and 3e). Furthermore, we assessed the overall mechanical contrast across the entire effective FOV of MSOP using whole-region segmentation masks defined by the automated boundary (black dashed curve in Fig. 3c). The continuous pixel-intensity distributions within the entire inclusion domain and the surrounding base matrix were compiled into a violin plot (Fig. 3f), mapping the clear mechanical separation achieved by the platform without human selection bias.

To characterise the strain-dependent mechanical behaviour across varying preloads (10%, 20% and 30%), we aggregated the performance metrics into comprehensive trend trajectories (Figs. 3g–i); the corresponding full-sequence image panels tracking 10% and 20% preloads are detailed in Supplementary Note 4 and Fig. S5. Quantitatively, the measured width of the inclusion exhibited a positive correlation with the applied preload (Fig. 3g), a phenomenon primarily attributed to the predictable lateral expansion (Poisson effect) of the silicone under axial compression. Notably, the inclusion maintained highly symmetrical lateral expansion across both the horizontal and vertical axes, confirming that the automated edge-detection framework remains robust during physical compression. Despite these expected morphological shifts under load, the stress contrast (ratio of the mean inclusion stress to the mean base stress, $\mu_{inclusion}/\mu_{base}$) increased monotonically with the preload (Fig. 3h). This behaviour is consistent with the non-uniform hardening of the silicone layer, where higher global strains amplify the intrinsic mechanical contrast between the stiff inclusion and the embedding compliant matrix. The platform's feature detectability was benchmarked using the global contrast-to-noise ratio (CNR, expressed as $\frac{\mu_{inclusion}-\mu_{base}}{\sigma_{base}}$, where $\sigma_{base}$ is the standard deviation of the base stress), with the Rose criterion (CNR = 4) defined as the conservative physical threshold for reliable feature detection (Fig. 3i). While the inclusion was robustly detectable even at a minimal 10% strain (CNR = 5.2), the signal quality improved significantly at the higher 30% preload, reaching a maximum CNR of 7.1 (Fig. 3i). These systematic trend lines confirm that the MSOP platform delivers an exceptional signal-to-noise ratio and maintains precise, sub-millimeter edge-localisation fidelity across varying operational strain regimes.

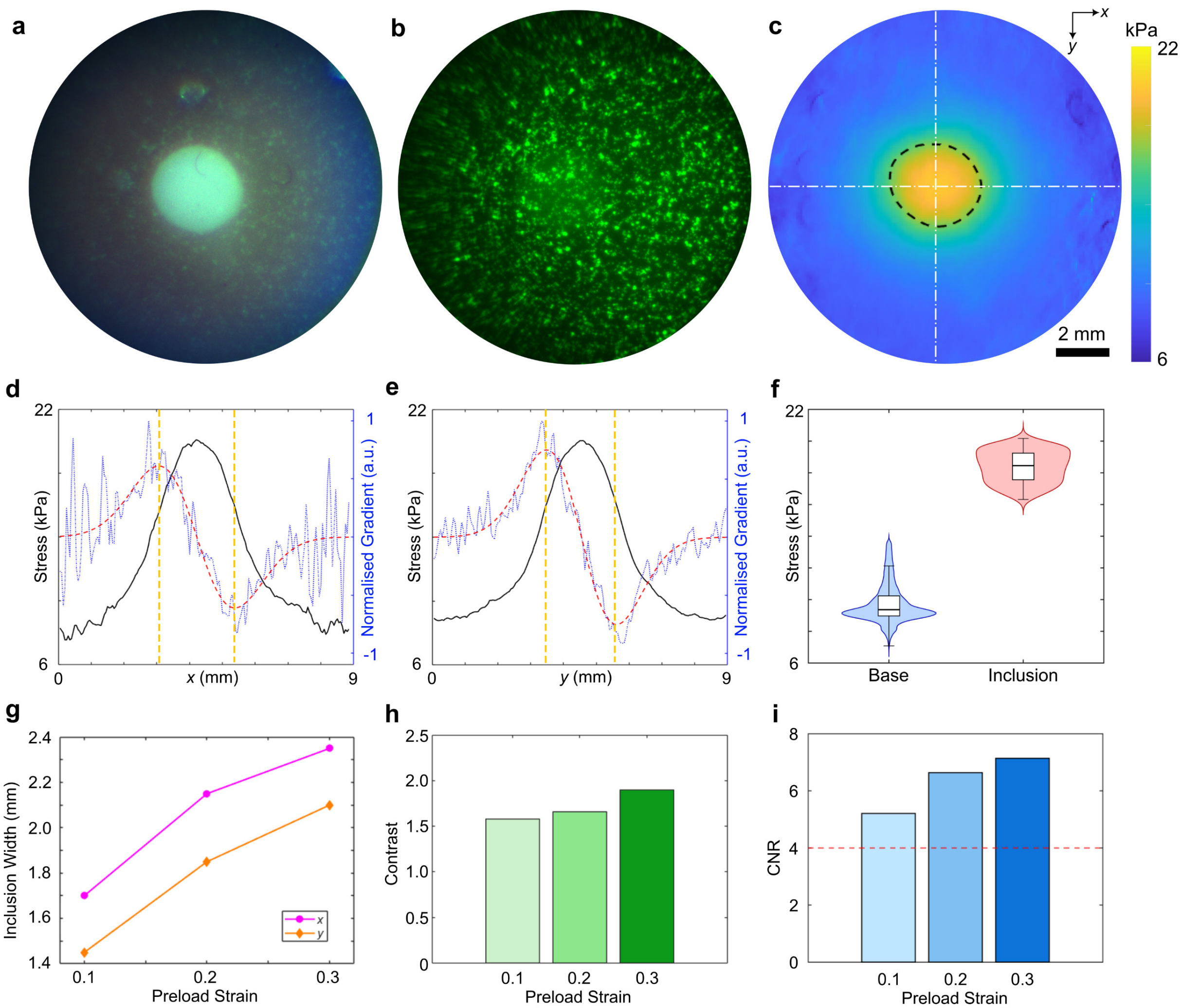

**Figure 3**. **Quantitative validation of MSOP and mechanical characterisation of silicone phantoms. a–c** White-light morphology, filtered UV-illuminated image, and the reconstructed stress map, respectively, of a silicone phantom with 2 mm inclusion. **d,e** Transverse stress profiles (black lines) along the horizontal and vertical axes (marked in **c**), with corresponding stress derivatives (blue dashed curves) and first-order Gaussian derivative fits (red dashed curves) used for edge localization (yellow dashed vertical lines). **f** Distribution of measured stress within the inclusion and base matrix (marked by black dashed curve in **c**), visualised via violin plots. The box-and-whisker elements represent the median, interquartile range, and minimum-to-maximum range. **g–i** Quantitative analysis of the 2 mm inclusion phantom as a function of preload strain, including measured inclusion width (**g**), stress contrast (**h**), and CNR (**i**). The red dashed line in (**i**) indicates the Rose criterion (CNR = 4).

## *Ex vivo* characterisation and detection of mouse pancreatic cancer

To evaluate the imaging and characterisation capability of MSOP in a biological context, we mapped the surface mechanical profiles of five fresh mouse pancreatic tissue specimens excised from a cancer model (Fig. 4). The experimental workflow, encompassing surgical excision and immediate *ex vivo* MSOP imaging, is illustrated in Fig. 4a. For each specimen, we captured sequential white-light morphological photographs and UV-illuminated images (Fig. 4b) to resolve the mechanical signatures of the tissue architecture. Accurate spatial correlation between MSOP stress maps and the histopathological "gold standard" is critical for validating tumour detection. To achieve this, we utilised large-scale morphological photography and corresponding histology as intermediate coordinate reference frames (Fig. 4c). Tissue locations evaluated by the MSOP system (marked by magenta dashed circles in Fig. 4c) were co-registered to the corresponding histology via landmark-based alignment, identifying common macroscale structural features across the large-scale photographs and the localised MSOP FOV.

The individual imaging and core mechanical profiles for two representative specimens are detailed in Figs. 4d and 4e, following a standardised sequence of analysis: (i) MSOP-captured white-light morphology, (ii) hematoxylin and eosin (H&E)-stained histology with malignant boundaries delineated by independent pathological assessment (black dashed curve), (iii) filtered UV-illuminated images, (iv) reconstructed stress maps with automatically detected tumour regions, (v) spatial overlays of the stress maps onto the white-light photographs, and (vi) continuous pixel-intensity stress distributions within the entire tumour and benign tissue regions. The comprehensive imaging pipelines for the remaining three specimens are compiled in Supplementary Note 5 and Fig. S6. Across all specimens, the MSOP system successfully delineated high-stress regions that closely corresponded with the histologically confirmed tumour boundaries. These mechanical boundaries [black dashed curves in panels (iv) and (v)] were automatically localised using the first-order Gaussian derivative edge-localisation algorithm established in our phantom validations. We observed minor spatial deviations between the reconstructed stress boundaries and the gold-standard histology, which are fundamentally attributed to well-documented *ex vivo* tissue warping, volumetric shrinkage, and dehydration that occur during the process of tissue embedding for histology. Despite these minor morphological shifts, the high-stress signatures remained highly localised to the malignant regions, whereas the surrounding benign tissue exhibited consistently lower and more uniform stress profiles.

Crucially, a comparative analysis between these two representative specimens highlights the distinct diagnostic advantage of MSOP over conventional optical inspection. In the first specimen (Fig. 4d), the mechanically delineated tumour boundary [black dashed curve in panel (iv)] corresponds to a surface region exhibiting distinct hypervascularisation and structural irregularity under white-light illumination [panel (i)], a macroscale feature driven by accelerated tumour angiogenesis and localised cell proliferation. Conversely, in the second specimen (Fig. 4e), the histologically verified malignant core is entirely indistinguishable from the surrounding benign tissue in the raw white-light photograph [panel (i)], yet it is clearly resolved as a high-contrast zone in the MSOP stress map [panel (iv)]. This capability underscores the clinical contribution of MSOP: by translating subsurface elasticity into a digital overlay, the platform provides an objective, contrast-enhanced diagnostic signature that unmasks concealed malignancies hidden to the naked eye, thereby overcoming the inherent limitations of standard visual inspection during oncological surgery.

To assess the robustness of MSOP across the entire animal cohort (n = 5), we compiled an aggregate summary of the mechanical metrics (Figs. 4f-h). The paired-dot plot (Fig. 4f) reveals a clear mechanical separation between the two tissue domains, where every specimen exhibited an elevation in mean stress within its automatically detected tumour region relative to the surrounding benign tissue. This discrimination is further highlighted by the stress contrast plot (Fig. 4g), which tracks the ratio of the mean tumour stress to mean benign tissue stress ($\mu_{tumour}/\mu_{benign}$) for each specimen, yielding a cohort-wide average contrast ratio of 1.9 (marked by the horizontal dashed line in Fig. 4g). Furthermore, we evaluated the system's sensitivity by computing the CNR ($\frac{\mu_{tumour}-\mu_{benign}}{\sigma_{benign}}$, where $\sigma_{benign}$ is the standard deviation of the benign tissue stress) across the cohort (Fig. 4h). The MSOP system achieved an average CNR of 3.7 (horizontal dashed line in Fig. 4h), demonstrating a signal-to-noise threshold that consistently enables automated tumour detection. Collectively, these results validate that MSOP provides highly reproducible mechanical contrasts capable of reliably differentiating malignant from benign pancreatic tissue, underscoring its translational potential as an objective label-free tool for intraoperative margin verification.

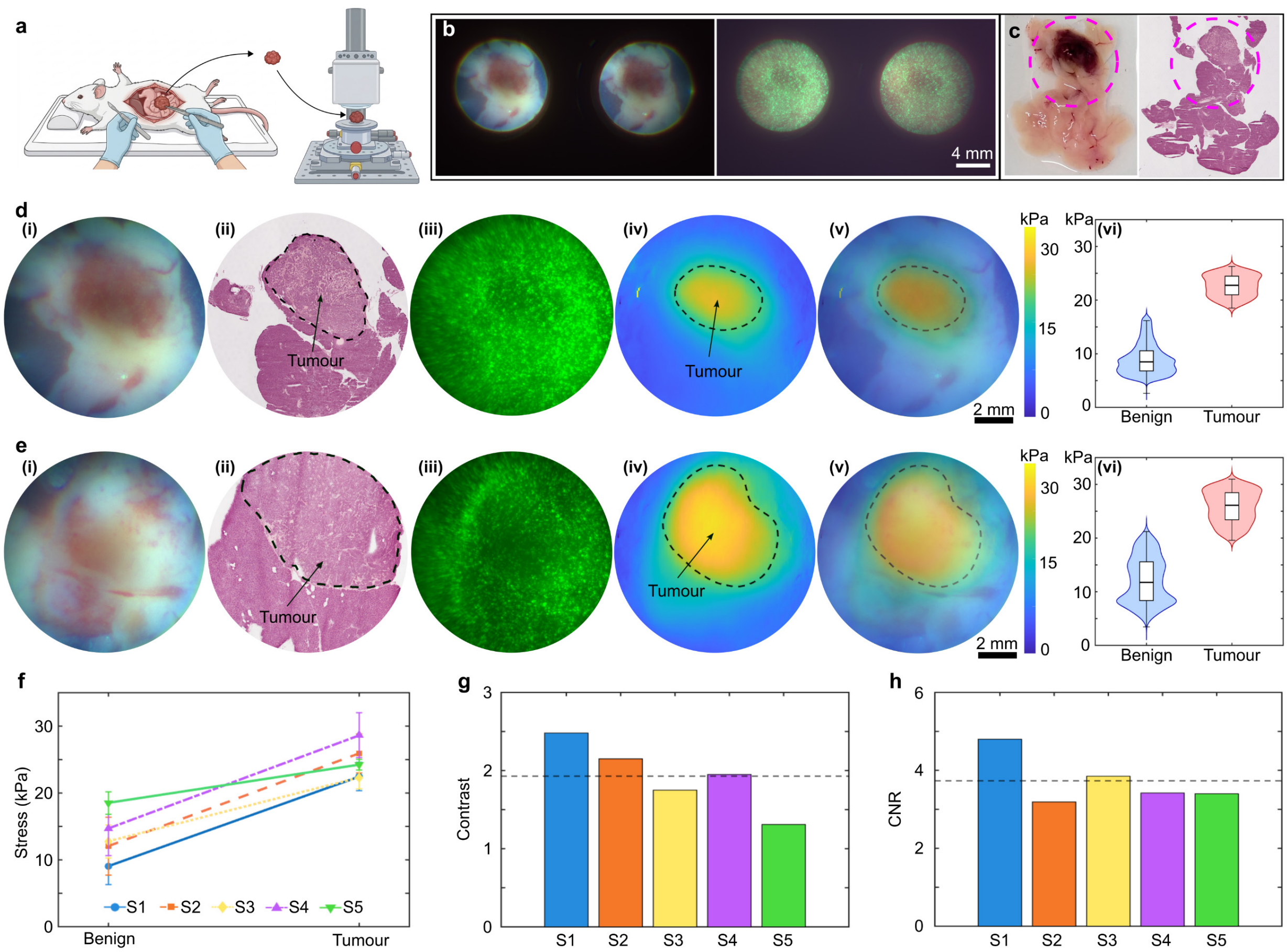

**Figure 4**. **MSOP imaging and quantitative characterisation of mouse pancreatic cancer. a** Schematic of the experimental pipeline, from surgical excision to immediate *ex vivo* MSOP imaging. **b** Representative raw white-light morphological photography and corresponding UV-illuminated images captured by the MSOP system. **c** Spatial co-registration workflow. Large-scale morphological photography (left) and H&E histology (right) serve as intermediate coordinate reference to align the MSOP FOV (magenta dashed circles) with the pathological gold standard. **d, e** Multi-modal imaging and mechanical characterisation of two representative mouse pancreatic tissue specimens. Sub-panels (**i**–**vi**) depict: (**i**) MSOP white-light morphology, (**ii**) H&E histology (black dashed curve indicates tumour boundary), (**iii**) filtered UV-illuminated images, (**iv**) reconstructed stress maps (black dashed curve indicates automatically detected boundary), (**v**) co-registered stress maps overlaid onto white-light photography, and (**vi**) continuous pixel-intensity stress distributions within the malignant and benign regions. **f** Paired-dot comparison of mean stress values between the automated tumour and benign tissue regions across the cohort; error bars denote the standard deviations of tissue stress within each region. **g** Stress contrast tracking individual specimens, with the cohort mean of 1.9 indicated by the horizontal dashed line. **h** CNR analysis across the cohort, with the cohort mean of 3.7 marked by the horizontal dashed line. For panels **f**-**h**, S1-S5 designate independent biological specimens 1-5, respectively.

## Clinical validation on human breast and liver cancer specimens

To demonstrate the translational potential of MSOP in clinical oncology, we mapped the surface mechanical profiles of freshly excised human tissue specimens obtained immediately following breast and liver cancer surgeries (Fig. 5). This clinical validation evaluates the system's capability to resolve mechanical heterogeneities within complex human tissue environments immediately post-excision without exogenous labeling. The imaging and analysis for both breast (Fig. 5a) and liver (Fig. 5b) specimens follow a comprehensive sequence: from surgical excision and immediate *ex vivo* MSOP imaging [panels (i) and (ii)], through multimodal and multi-scale histopathological co-registration and quantitative mechanical mapping [panels (iii)–(viii)], to whole-region stress distribution and CNR analyses [panels (ix) and (x)].

Spatial co-registration between the stress maps and the histopathological "gold standard" was achieved by aligning the localised MSOP FOV with large-scale morphological photography and structural landmarks on the subsequent H&E-stained slides. Detailed pathological assessment of the zoomed-in histology [panel (vi)] served as the structural ground truth for identifying malignant regions. In both organs, the reconstructed MSOP stress maps [panel (vii)] and the resulting stress-morphology overlays [panel (viii)] demonstrated close spatial correspondence with the histologically confirmed tumour boundaries. These mechanical margins were automatically delineated using the first-order Gaussian derivative edge-localisation algorithm established in our phantom validations. Minor morphological discrepancies between the mechanical boundaries and the gold-standard histology are fundamentally attributed to non-uniform tissue warping, volumetric shrinkage, and dehydration typical of the formalin-fixation and paraffin-embedding (FFPE) process.

To rigorously evaluate the system's boundary-tracking fidelity and ensure that the reconstructed mechanical contrast remains stable across different spatial scales, we implemented an analytical framework combining whole-region global segmentation with localised, boundary-relative proximity testing [panels (vii) and (viii)]. The global tumour boundaries were determined purely by the automated edge-localisation algorithm [black dashed curves in panels (vii) and (viii)], encompassing the entire continuous pixel population within the FOV. Concurrently, to evaluate margin precision and confirm that the mechanical signal is uniformly distributed along the interface rather than skewed by isolated hotspots, we extracted three paired tumour/benign regions (1 mm diameter) positioned systematically at a fixed distance of 1 mm normal to the automatically defined tumour boundary [labeled 1-3 in panel (viii)].

Quantitative comparison between the segmented tissue domains revealed distinct mechanical signatures characteristic of each organ's pathological landscape. While both clinical specimens exhibited clear tumour-to-benign contrast [panels (ix)], the global stress contrast ($\mu_{tumour}/\mu_{benign}$) was recorded at 1.4 for the breast tissue and 2.9 for the liver tissue. This variance in mechanical mismatch reflects the distinct structural environments of the two malignancies. Specifically, the extensive desmoplasia and intense fibrotic remodeling characteristic of hepatocellular carcinoma (HCC) induce a more profound mechanical mismatch against the compliant surrounding liver parenchyma than the invasive ductal stiffening typically observed within the adipose-rich stroma of the breast.

This structural variance directly governed the signal discriminability of the tumour boundaries [panels (x)]. For the highly contrasting liver specimen, the global tissue-wide matrix yielded a definitive CNR of 4.8, comfortably clearing the Rose criterion (CNR = 4) for unambiguous feature detection. Conversely, for the breast specimen, the global tissue-wide CNR converged immediately at this fundamental physical threshold, reaching 3.9. Crucially, however, the localised boundary-relative ROI pairs consistently exceeded the Rose criterion across both tissue types [panels (x)], despite being positioned a mere 1 mm from the automated boundary. Achieving such definitive signal separation at this spatial proximity demonstrates that MSOP can delineate sharp elastographic gradient, successfully resisting the mechanical blurring that typically obscures transition zones. Taken together, these human pilot data demonstrate that MSOP serves as a versatile, label-free imaging modality capable of mapping diverse human malignancies and delineating margins based entirely on their intrinsic mechanical signatures.

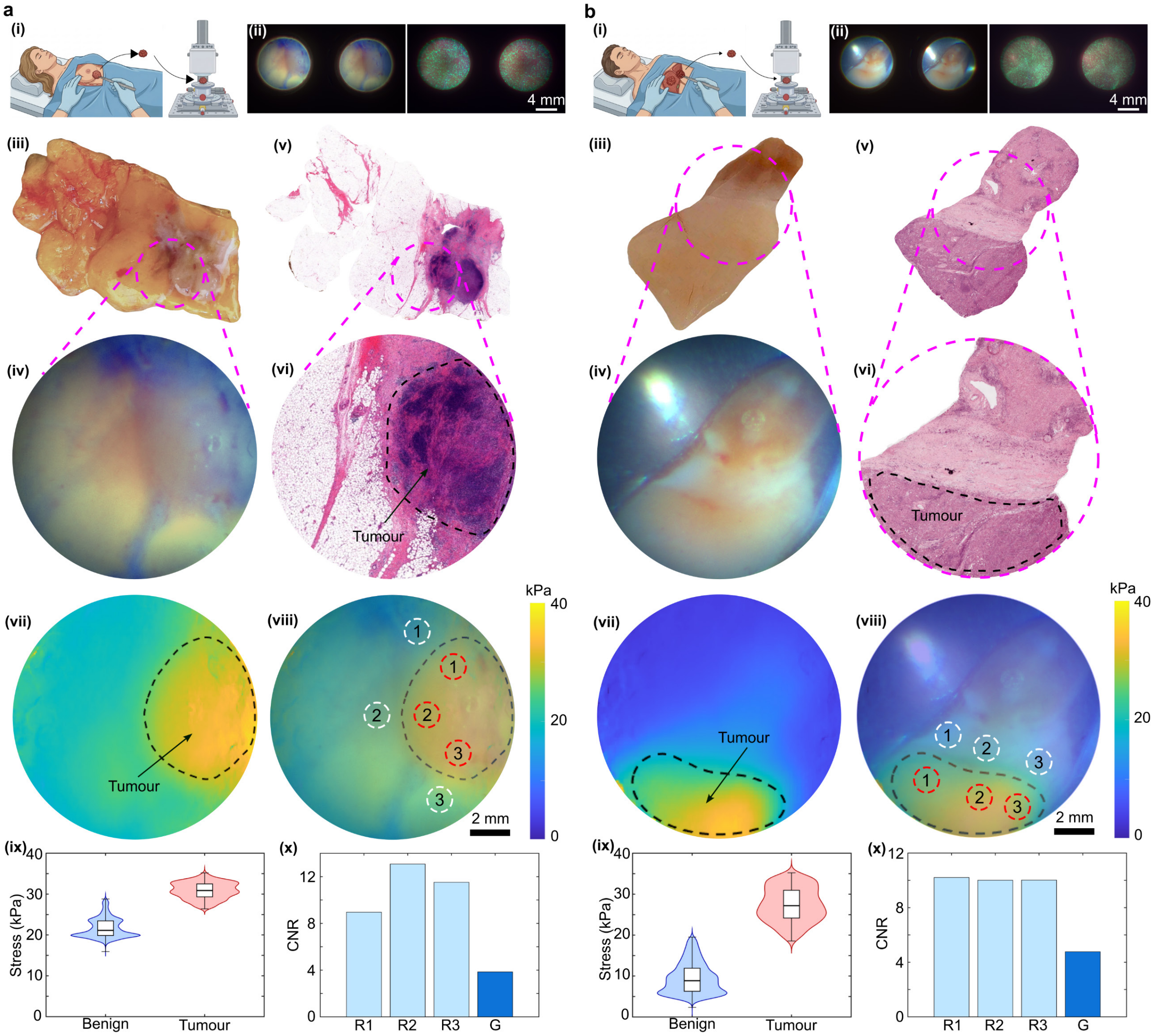

**Figure 5**. **MSOP imaging of human breast and liver cancer specimens. a, b** Comprehensive mechanical characterisation of freshly excised human breast cancer (**a**) and liver cancer (**b**) tissue specimens. Sub-panels follow a consistent analysis sequence: (i) schematic of the imaging workflow; (ii) raw white-light and UV-illuminated images; (iii) large-scale morphological reference photography; (iv) MSOP-captured morphology aligned via landmark co-registration; (v, vi) whole-mount and zoomed-in H&E histology with tumour regions (black dashed curves) delineated by pathological assessment; (vii, viii) reconstructed stress maps and stress/white-light overlays, with detected tumour boundaries marked by black dashed curves and ROIs marked by small dashed circles; (ix) stress distribution for the entire tumour (red) and benign tissue (blue) regions; and (x) CNR analysis for the three ROI pairs and the global tumour regions. R1-R3, ROI 1-3. G, global regions.

## Discussion

In this study, we established MSOP as a compact, single-sensor platform for the high-contrast mechanical mapping of biological tissues for tumour margin assessment. Our results across heterogeneous phantoms, mouse models, and human clinical specimens confirm that MSOP can resolve sub-millimeter mechanical heterogeneities, providing a quantitative, digital alternative to the subjective manual palpation.

Previous dual-camera stereoscopic systems require complex extrinsic calibration and high-speed hardware triggering to align discrete optical paths, processes that are prone to mechanical drift and temporal lag in dynamic surgical environments. By multiplexing binocular perspectives onto a single CMOS sensor through an integrated, flat-optic metalens, our imaging platform eliminates inter-sensor calibration errors and spatial misalignment. This architecture ensures that the stereoscopic views used for stress reconstruction are inherently locked both spatially and temporally, resulting in a robust, drift-free transduction of optical displacement into quantitative mechanical contrast. To our knowledge, this work represents the first application of a binocular metalens to a functional biomedical diagnostic task. By successfully bridging the gap between flat optics and optical elastography, these results demonstrate that metalens can surpass purely structural imaging and serve as a practical, translational platform to enhance the clinical usability of real-time surgical tissue characterisation.

Furthermore, the integrated, 3D-printed design of the MSOP imaging head (encapsulating the CMOS sensor, metalens, and illumination channels within a unified chassis) highlights the inherent capacity for miniaturisation. While the current prototype serves as a standalone proof-of-principle platform, the planar, chip-scale nature of the meta-optics allows for further reduction in the optical design. This scalability paves the way for integrating MSOP into ultra-compact handheld surgical probes or even finger-mounted tactile sensors. Such form factors would substantially reduce the physical envelope compared to previously reported handheld elastography probes (41-43), ultimately providing surgeons with an "augmented touch" capability to map tumour margins during both open and minimally invasive procedures.

To fully realise this clinical translation, however, several engineering steps must be addressed. While the current prototype footprint (6×6×6 $cm^3$) represents a greater than tenfold reduction in volume compared to benchtop dual-camera setups (12×12×20 $cm^3$), further scaling is required for intraoperative *in vivo* deployment, particularly for inclusion within the restricted form factors of minimally invasive or robotic surgical devices. Future iterations will leverage advanced nanofabrication to integrate the metalens directly onto the CMOS cover glass. Additionally, our current post-processing pipeline relies on offline DIC calculations; achieving real-time intraoperative feedback will require the implementation of GPU-accelerated algorithms capable of live stress-map rendering. While our validation on human breast and liver specimens provided high contrast, future large-scale clinical trials with increased sample sizes are necessary to establish the diagnostic sensitivity and specificity required for regulatory approval.

In summary, we have developed and validated a stereoscopic metalens-based platform that digitises the mechanical contrast of biological tissues. By merging the precision of flat-optic wavefront engineering with the principles of optical palpation, MSOP provides a label-free, high-contrast imaging tool capable of delineating solid tumour boundaries. As surgical oncology moves toward increasingly precise, minimally invasive interventions, highly integrated meta-optical systems like MSOP will play a pivotal role in ensuring complete tumour resection and improving long-term patient outcomes. Furthermore, the capacity of this compact architecture to resolve spatiotemporal surface profiles at a mesoscopic scale opens a clear translational pathway for versatile applications beyond oncology, ranging from real-time haptic feedback in robotic surgical tools to localised cardiovascular strain mapping in dynamic physiological environments.

# Methods

## Metalens numerical simulations

For the optical palpation system, a binocular metalens is designed and fabricated at a target wavelength 550 nm. The amplitude and phase of the transmitted light through the circular $Si_3N_4$ nanopillars are calculated using a MATLAB-based rigorous coupled-wave analysis (RCWA) solver. For each candidate pillar radius r (fixed height h = 750 nm, period p = 350 nm), the transmission at 550 nm is obtained (Supplementary Note 1, Fig. S1), and a subset of radii is selected to provide continuous $0 - 2\pi$ phase coverage while maintaining high transmittance. The required lens phase $\phi_{lens}(x, y)$is then discretized by assigning to each lattice site the radius whose $\phi(r)$most closely matches $\phi_{lens}(x, y)$, yielding a uniform-height, radius-encoded phase profile suitable for fabrication. To check the performance of the designed metalens, the resulting complex pupil field, which is constructed from the RCWA-predicted amplitude and the discretised phase, is propagated using the angular spectrum method to predict the axial intensity distribution, from which the focal distance is identified at the on-axis maximum (Fig. 2c).

## Metalens fabrication

Fabrication of the binocular metalens was carried out by depositing a 750 nm $Si_3N_4$ layer on a 500 µm quartz substrate using plasma-enhanced chemical vapour deposition (Oxford PlasmaPro100 PECVD) at 200℃, at a deposition rate of 0.5 nm/min in a mixture of $SiH_4$, $NH_3$ and $N_2$ gases. After the deposition, PMMA (A6) e-beam resist was spin-coated onto the $Si_3N_4$ film and baked at 180 °C for 180 seconds. To increase the surface conductivity during exposure, a water-soluble conductive polymer (Allresist, Electra 92) was spin-coated and baked at 90°C for 60 seconds. Pattern definition was performed on a 100-kV electron-beam lithography system (Raith, EBPG 5000). Prior to development, the conductive polymer was removed by a DI-water rinse and the sample was dried with nitrogen. The resist was then developed in MIBK/IPA (1:3) mixture for 60 seconds to reveal the pattern. Following this, a 30-nm thick Cr layer was deposited by E-beam evaporation (Angstrom, Amod). Lift-off was carried out in acetone to remove resist and Cr from the unexposed regions, leaving the Cr hard mask pattern on the substrate. The patterns were subsequently transferred into the $Si_3N_4$ layer by deep reactive-ion etching (Oxford PlasmaPro 100, Estrelas) using mixed $CHF_3/O_2$ gases. An $O_2$ plasma clean was applied for 3 min to remove residual organics. Finally, the Cr hard mask was removed by wet etching (commercial Cr etchant), yielding the completed metalens.

## Metalens optical characterisation

To characterise the performance of the single metalens, a custom optical system was assembled. The full layout of the characterisation setup is provided in Supplementary Note 2. Axial intensity profile images were acquired by a CMOS camera while translating the sample in increments of 300nm using a Thorlab linear translation stage (Thorlab, DDS300/M) with Direct-drive brushless servo motor (Thorlab, BBD303) under Python-script control.

## Digital image correlation and disparity calculation

The stereoscopic disparity maps were computed using Ncorr, an open-source two-dimensional DIC framework implemented in MATLAB. Prior to correlation, the raw UV-illuminated stereoscopic pairs were spectrally filtered (green channel) to isolate the phosphorescent microparticle emissions and enhance the contrast of the speckle pattern. The left perspective was designated as the reference image, while the right perspective served as the current image. A circular subset with a diameter of 500 µm was employed to track local intensity features across the FOV. The DIC algorithm utilised a nonlinear least-squares solver to minimise the cross-correlation cost function, with a convergence criterion set at a displacement threshold of $10^{-4}$. For each correlated feature, the horizontal shift in the $x$-coordinate between the reference and current images was extracted as the local optical disparity. To ensure high processing speed and data continuity, a step size of 6 pixels (5-pixel spacing, 50 μm) was used for the displacement grid.

## Silicone layer and phantom fabrication

In SOP, to enable sequential acquisition of stress maps and white-light photographs at the same tissue location, we developed a silicone layer containing phosphorescent microparticles which appear transparent under white light illumination and emit green light under UV light illumination (24). The phosphorescent microparticles (Glowing Gecko, diameter ≤30 μm) were randomly distributed at the interface between the two layer sections that comprised the layer (5 mm thick top section and 0.2 mm thick bottom section, Wacker Elastosil P7676 A and B at 1:1 mixing ratio, elasticity = 30.7 ± 1.9 kPa at 20% strain). The two layer sections were then cured together to achieve a total thickness of 5.2 mm and a diameter of 20 mm.

The inclusion phantom consists of a soft bulk (Smooth-On Ecoflex 00-20 A and B at 1:1 mixing ratio, elasticity = 36.9 ± 0.8 kPa at 20% strain, diameter = 20 mm, thickness = 5 mm) and a stiff cylindrical inclusion (Wacker Elastosil RT601 A and B and Wacker AK 50 at 10 : 1 : 10 mixing ratio, elasticity = 225.6 ± 11.2 kPa at 20% strain, thickness = 4.5 mm). The inclusion was embedded at the centre of the phantom, 200 μm below the surface. Titanium dioxide particles were incorporated in the inclusion at a mixing ratio of 2mg/ml to provide visual contrast to the transparent bulk.

## Mouse model and tissue preparation

A genetic pancreatic cancer model was used in this study. Transgenic RIP1-Tag5 mice were bred on a C3HeBFe (C3H) background, as previously described (44, 45). All mice were maintained under pathogen-free conditions at the University of Western Australia with food and water provided ad libitum. All animal studies were approved by the UWA Animal Ethics Committee (protocols ET0000455 and ET0000492). RIP1-Tag5 mice develop pancreatic neuroendocrine tumours between 23 and 30 weeks of age; whole pancreas containing tumours were collected from 27-week-old mice. Fresh pancreatic tissue specimens were immediately characterised using the MSOP system, followed by embedding in optimal cutting temperature compound while preserving the same tissue orientation. Cryosections (7 μm thickness) were stained with H&E according to standard protocols. High-resolution digital micrographs of the stained sections were reviewed by a specialist to delineate malignant and benign regions, providing the structural ground truth for spatial co-registration with the reconstructed mechanical stress maps.

## Human tissue specimens and histopathological preparation

We performed MSOP on a breast and a liver tissue specimen, freshly excised from patients undergoing surgery at Fiona Stanley Hospital, Western Australia with informed consent. The breast tissue ethics was approved by the South Metropolitan Health Service Human Research Ethics Committee (PRN: RGS0000003726), and the liver tissue was obtained from the Perkins Translational Cancer Biobank under human ethics approval from the WA Health Central Human Research Ethics Committee (PRN: RGS0000000919). The breast specimen was excised from a patient diagnosed with invasive ductal carcinoma (IDC) and undergoing breast mastectomy. The liver specimen originated from a surgical resection procedure in a patient diagnosed with HCC and enrolled in the Liver Cancer Collaborative (LCC) cohort. Each MSOP image was co-registered with post-operative histology acquired ∼1 week after the imaging. The histology of breast tissue was performed at PathWest Laboratory Medicine at Fiona Stanley Hospital, and the liver tissue histology was performed at Harry Perkins Institute of Medical Research.

Liver Cancer Collaborative:

Peter J Leedman[6] Janina E. E. Tirnitz-Parker[15], Michael C. Wallace[16], Louise N. Winteringham[6],

*[15]Curtin Medical Research Institute and Curtin Medical School, Curtin University, Perth, WA, Australia.*

[16]*Department of Hepatology, Sir Charles Gairdner Hospital, Perth, WA, Australia.*

Acknowledgements

Q.F.: WANMA Emerging Leaders Fellowship, Department of Health, Government of Western Australia; Safe Harbour Awards, Harry Perkins Institute of Medical Research.

H.R.: Australian Research Council grants (DP220102152, FT250100565);

S.A.M.: Australian Research Council grants (DP220102152); Lee Lucas Chair in Physics;

This work was performed in part at the Melbourne Centre for Nanofabrication (MCN) in the Victorian Node of the Australian National Fabrication Facility (ANFF).

Author contributions

Q.F. and H.R. conceived and designed the project. H.Y., R.J., C.L., H.R., and Q.F. built the system hardware. H.Y., C.L., and H.R. performed the metalens experiments and data analysis. R.J. and Q.F. performed the MSOP experiments. Q.F. developed the edge-detection algorithm, and analysed and interpreted the data. B.H. prepared the animal models and performed the animal and human liver histology. R.Z. and L.N.W. coordinated the human tissue acquisition. L.G. and M.H. performed the human breast histology. F.B., S.H., and C.M.S. recruited the patients. Q.F. and H.Y. wrote the manuscript with input from all authors. Q.F., H.R., B.F.K., and S.A.M. supervised the project.

Data availability

The main data supporting the results in this study are available within the paper and its Supplementary Information. Other data are too large to be publicly shared, yet they are available for research purposes from the corresponding authors on reasonable request.

Competing interests

B.F.K. and C.M.S. have a financial interest in OncoRes Medical Pty Ltd. This company did not provide support for this work. All other authors declare no competing interests.

## References:

1. Bray F, Laversanne M, Sung H, Ferlay J, Siegel RL, Soerjomataram I, et al. Global cancer statistics 2022: GLOBOCAN estimates of incidence and mortality worldwide for 36 cancers in 185 countries. CA Cancer J Clin. 2024;74(3):229-63.
2. Sullivan R, Alatise OI, Anderson BO, Audisio R, Autier P, Aggarwal A, et al. Global cancer surgery: delivering safe, affordable, and timely cancer surgery. The Lancet Oncology. 2015;16(11):1193-224.
3. Wyld L, Audisio RA, Poston GJ. The evolution of cancer surgery and future perspectives. Nature Reviews Clinical Oncology. 2015;12(2):115-24.
4. Gorman BG, Hanson J, Vidal NY. The importance of palpation in the skin cancer screening examination. Journal of Cosmetic Dermatology. 2021;20(12):3982-5.
5. Macherey S, Doerr F, Heldwein M, Hekmat K. Is manual palpation of the lung necessary in patients undergoing pulmonary metastasectomy? Interactive CardioVascular and Thoracic Surgery. 2015;22(3):351-9.
6. Konstantinova J, Li M, Mehra G, Dasgupta P, Althoefer K, Nanayakkara T. Behavioral Characteristics of Manual Palpation to Localize Hard Nodules in Soft Tissues. IEEE Transactions on Biomedical Engineering. 2014;61(6):1651-9.
7. Chaturvedi P, Datta S, Nair S, Nair D, Pawar P, Vaishampayan S, et al. Gross examination by the surgeon as an alternative to frozen section for assessment of adequacy of surgical margin in head and neck squamous cell carcinoma. Head & Neck. 2014;36(4):557-63.
8. Yenigün BM, Yüksel C, Kahya Y, Görgüner F, Çoruh Gürsoy A, Kocaman G, et al. Effectiveness of intraoperative bimanual palpation in metastatic tumors of lung. Turk Gogus Kalp Damar Cerrahisi Derg. 2020;28(4):662-8.
9. Zhang B, Zhang Y, Le H, Li W, Chen C, Fang R, et al. Intraoperative localization in minimally invasive surgery for small pulmonary nodules: a retrospective study. Transl Cancer Res. 2021;10(7):3470-8.
10. Hari S, Kumari S, Srivastava A, Thulkar S, Mathur S, Veedu PT. Image guided versus palpation guided core needle biopsy of palpable breast masses: a prospective study. Indian J Med Res. 2016;143(5):597-604.
11. Nakajima J, Murakawa T, Fukami T, Sano A, Sugiura M, Takamoto S. Is finger palpation at operation indispensable for pulmonary metastasectomy in colorectal cancer? Ann Thorac Surg. 2007;84(5):1680-4.
12. Yamaoka Y, Kinugasa Y, Shiomi A, Yamaguchi T, Kagawa H, Yamakawa Y, et al. Is it important to palpate lymph nodes in open surgery for colorectal cancer? Asian J Endosc Surg. 2017;10(2):143-7.
13. Watkinson JC, Johnston D, Jones N, Coady M, Laws D, Allen S, et al. The reliability of palpation in the assessment of tumours. Clin Otolaryngol Allied Sci. 1990;15(5):405-9.
14. Kennedy BF, Wijesinghe P, Sampson DD. The emergence of optical elastography in biomedicine. Nature photonics. 2017;11:215-21.
15. Allen WM, Wijesinghe P, Dessauvagie BF, Latham B, Saunders CM, Kennedy BF. Optical palpation for the visualization of tumor in human breast tissue. Journal of biophotonics. 2019;12(1):e201800180.
16. Jones R, Fang Q, Kennedy BF. Analysis of image formation in optical palpation. Journal of biophotonics. 2024;n/a(n/a):e202400180.
17. Othman W, Lai Z-HA, Abril C, Barajas-Gamboa JS, Corcelles R, Kroh M, et al. Tactile Sensing for Minimally Invasive Surgery: Conventional Methods and Potential Emerging Tactile Technologies. Frontiers in Robotics and AI. 2022;Volume 8 - 2021.
18. Bandari N, Dargahi J, Packirisamy M. Tactile Sensors for Minimally Invasive Surgery: A Review of the State-of-the-Art, Applications, and Perspectives. IEEE Access. 2020;8:7682-708.
19. Naidu AS, Patel RV, Naish MD. Low-Cost Disposable Tactile Sensors for Palpation in Minimally Invasive Surgery. IEEE/ASME Transactions on Mechatronics. 2017;22(1):127-37.
20. Konstantinova J, Jiang A, Althoefer K, Dasgupta P, Nanayakkara T. Implementation of Tactile Sensing for Palpation in Robot-Assisted Minimally Invasive Surgery: A Review. IEEE Sensors Journal. 2014;14(8):2490-501.
21. Kennedy BF. Optical Coherence Elastography: AIP Publishing LLC; 2021. 384 p.
22. Fang Q, Wijesinghe P, Jones R, Lakhiani DD, Dessauvagie BF, Latham B, et al. Coherence function-encoded optical palpation. Optics letters. 2021;46(18):4534-7.

23. Sanderson RW, Fang Q, Curatolo A, Adams W, Lakhiani DD, Ismail HM, et al. Camera-based optical palpation. Scientific reports. 2020;10(1):15951.
24. Fang Q, Choi S, Taba A, Lakhiani DD, Newman K, Zilkens R, et al. Stereoscopic optical palpation for tumour margin assessment in breast-conserving surgery. Optics and Lasers in Engineering. 2023;166:107582.
25. Fang Q, Sanderson RW, Zilkens R, Boman I, Foo KY, Lakhiani DD, et al. Diagnostic feasibility study of stereoscopic optical palpation for breast tumour margin assessment. BMC Cancer. 2025;25(1):1793.
26. Yu N, Genevet P, Kats MA, Aieta F, Tetienne J-P, Capasso F, et al. Light Propagation with Phase Discontinuities: Generalized Laws of Reflection and Refraction. Science. 2011;334(6054):333-7.
27. Chen H-T, Taylor AJ, Yu N. A review of metasurfaces: physics and applications. Reports on Progress in Physics. 2016;79(7):076401.
28. Wang L, Kruk S, Koshelev K, Kravchenko I, Luther-Davies B, Kivshar Y. Nonlinear Wavefront Control with All-Dielectric Metasurfaces. Nano Letters. 2018;18(6):3978-84.
29. Khorasaninejad M, Chen WT, Devlin RC, Oh J, Zhu AY, Capasso F. Metalenses at visible wavelengths: Diffraction-limited focusing and subwavelength resolution imaging. Science. 2016;352(6290):1190-4.
30. Wang S, Wu PC, Su V-C, Lai Y-C, Chen M-K, Kuo HY, et al. A broadband achromatic metalens in the visible. Nature Nanotechnology. 2018;13(3):227-32.
31. Chen WT, Zhu AY, Sanjeev V, Khorasaninejad M, Shi Z, Lee E, et al. A broadband achromatic metalens for focusing and imaging in the visible. Nature Nanotechnology. 2018;13(3):220-6.
32. Ren H, Jang J, Li C, Aigner A, Plidschun M, Kim J, et al. An achromatic metafiber for focusing and imaging across the entire telecommunication range. Nature communications. 2022;13(1):4183.
33. Liu X, Chen MK, Chu CH, Zhang J, Leng B, Yamaguchi T, et al. Underwater Binocular Meta-lens. ACS Photonics. 2023;10(7):2382-9.
34. Zhang L, Yang J, Zhang L, Jing X, Liang C, Zhang C, et al. Four-dimensional imaging based on a binocular chiral metalens. Optics letters. 2025;50(3):1017-20.
35. Song Y, Zhang Y, Liu X, Tanaka T, Chen MK, Geng Z. High-precision three-dimensional imaging based on binocular meta-lens and optical clue fusion. npj Nanophotonics. 2025;2(1):28.
36. Zhao Z, Liu X, Ji Y, Zhang Y, Chen Y, Luo Z, et al. Meta-lens digital image correlation. Opto-Electronic Advances. 2025;8(9):250014-1--12.
37. Pahlevaninezhad H, Khorasaninejad M, Huang Y-W, Shi Z, Hariri LP, Adams DC, et al. Nano-optic endoscope for high-resolution optical coherence tomography in vivo. Nature photonics. 2018;12(9):540-7.
38. Davies S, Hu Y, Jiang N, Blyth J, Kaminska M, Liu Y, et al. Holographic Sensors in Biotechnology. Advanced Functional Materials. 2021;31(47):2105645.
39. Zhang S, Wong CL, Zeng S, Bi R, Tai K, Dholakia K, et al. Metasurfaces for biomedical applications: imaging and sensing from a nanophotonics perspective. Nanophotonics. 2021;10(1):259-93.
40. Born M, Wolf E. Principles of Optics: Electromagnetic Theory of Propagation, Interference and Diffraction of Light. 7 ed. Cambridge: Cambridge University Press; 1999.
41. Fang Q, Krajancich B, Chin L, Zilkens R, Curatolo A, Frewer L, et al. Handheld probe for quantitative micro-elastography. Biomedical Optics Express. 2019;10(8):4034-49.
42. Gong P, Chin SL, Allen WM, Ballal H, Anstie JD, Chin L, et al. Quantitative Micro-Elastography Enables In Vivo Detection of Residual Cancer in the Surgical Cavity during Breast-Conserving Surgery. Cancer research. 2022;82(21):4093-104.
43. Jones R, Zilkens R, Bharakhda A, Hardie M, Saunders CM, Fang Q, et al. A wireless and handheld optical palpation imaging probe for use in breast-conserving surgery. APL Bioeng. 2026;10(2):026104.
44. Li ZJ, He B, Domenichini A, Satiaputra J, Wood KH, Lakhiani DD, et al. Pericyte phenotype switching alleviates immunosuppression and sensitizes vascularized tumors to immunotherapy in preclinical models. J Clin Invest. 2024;134(18).
45. He B, Wood KH, Li ZJ, Ermer JA, Li J, Bastow ER, et al. Selective tubulin-binding drugs induce pericyte phenotype switching and anti-cancer immunity. EMBO Mol Med. 2025;17(5):1071-100.

# Supplementary

## Supplementary Note 1: Phase Library Construction and Unit-Cell Simulation

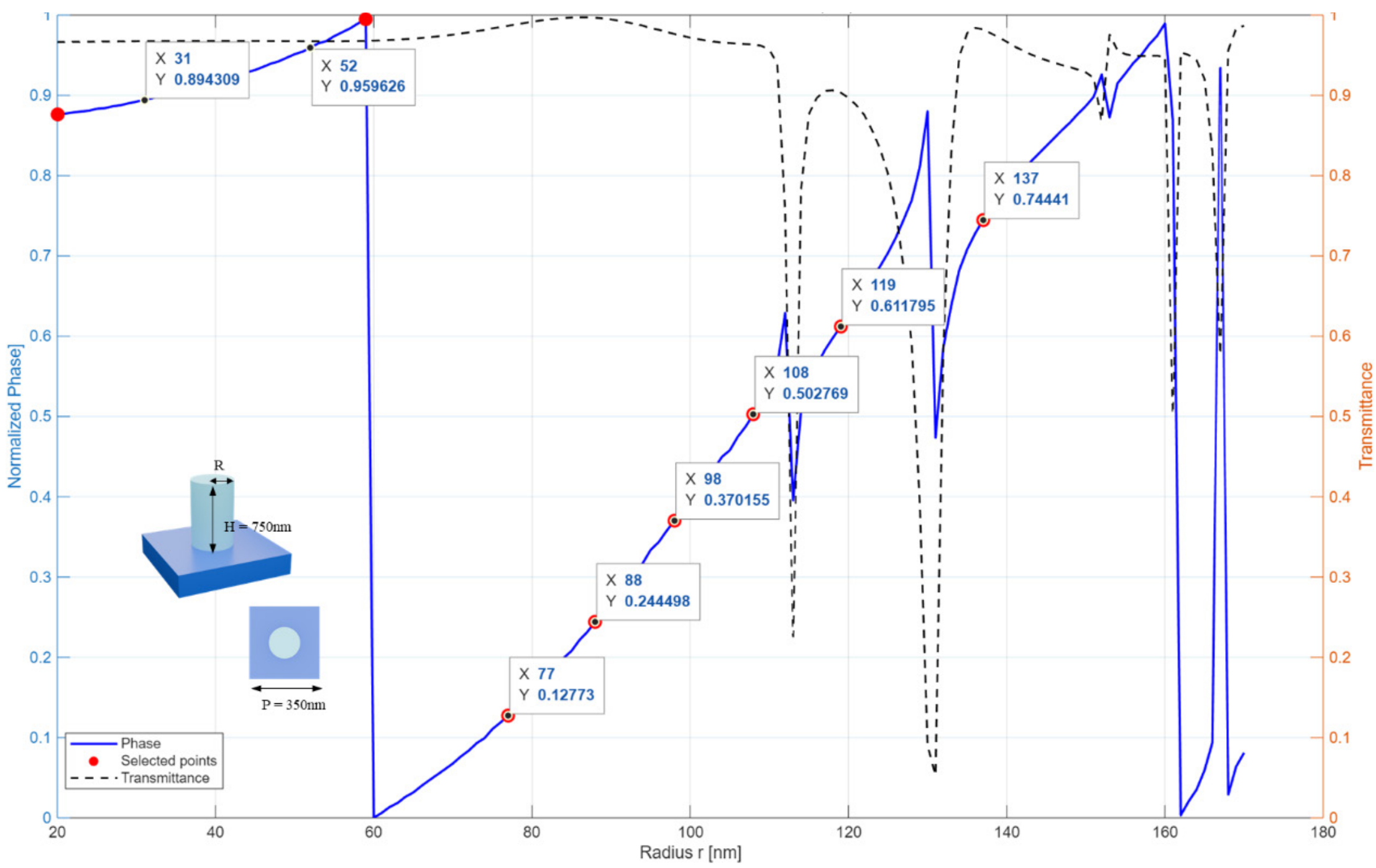

**Figure S1. Phase and transmittance of a circular Si3N4 nanopillar versus radius at λ = 550 nm, with pillar height H = 750 nm and period = 350 nm fixed.** The blue curve shows the transmitted phase (normalized to ([0,1]) ≡ ([0, $2\pi$])) as the pillar radius R is varied, while the black dashed curve (right axis) gives the corresponding power transmittance. Labels report ($R$ [$nm$], $normalized\ phase$). Insets illustrate the unit cell ($Si_3N_4$ pillar with height H = 750nm on a quartz substrate) and the top view indicating the period (P=350 nm). Red markers indicate the automatically selected candidates from an eight-bin phase segmentation, rather than the final radii used in fabrication.

Figure S1 shows the sweep of radius (R) of a cylindrical $Si_3N_4$ pillar (refractive index (n= 2.0) at 550 nm) on a quartz substrate under normal incidence. The pillar height was fixed at (H=750 nm) and the lattice period at (P=350 nm). For each (R), the complex 0th-order transmission coefficient was computed via RCWA to extract the transmitted phase and amplitude.

An initial set of candidate radii was automatically produced by dividing the target phase range (0)–($2\pi$) into eight equal bins and, within each bin, selecting a radius that satisfied (T>0.9) while best matching the bin-center phase. Because sudden phase wraps (typically near (R ~ 60 – 160 nm) coincide with resonant features where the amplitude may dip, design points were preferentially taken from monotonic, high-throughput segments to ensure uniform efficiency.

In view of fabrication tolerances, radii whose diameter approaches the period were avoided, and regions with locally smooth transmittance versus (R) were favored. The manually annotated values in the figure indicate the final radii actually used in the mask layout after tolerance screening. The resulting lookup table ($R_i, \phi_i$) provides a one-to-one radius–phase mapping for mask generation.

## Supplementary Note 2: Experimental setup for characterizing the performances of broadband achromatic metalens and array

To evaluate the metalens, we built a compact optical setup (Fig. Sxx). A supercontinuum source (NKT Photonics, SuperK FIANIUM) provided broadband collimated illumination. After wavelength selection with a band-pass filter centered at 550 nm, the beam passed through the substrate and was focused by the metalens to a diffraction-limited spot. The focused light was collected by a 50× long-working-distance objective (NA = 0.42) and relayed to a camera with a 200 mm commercial lens. The objective and relay lens were arranged confocally. The sample, mounted on a 3-axis stage, was scanned along z in 100 nm steps using a Thorlabs linear translation stage (DDS300/M) driven by a direct-drive brushless controller (BBD303) under Python control.

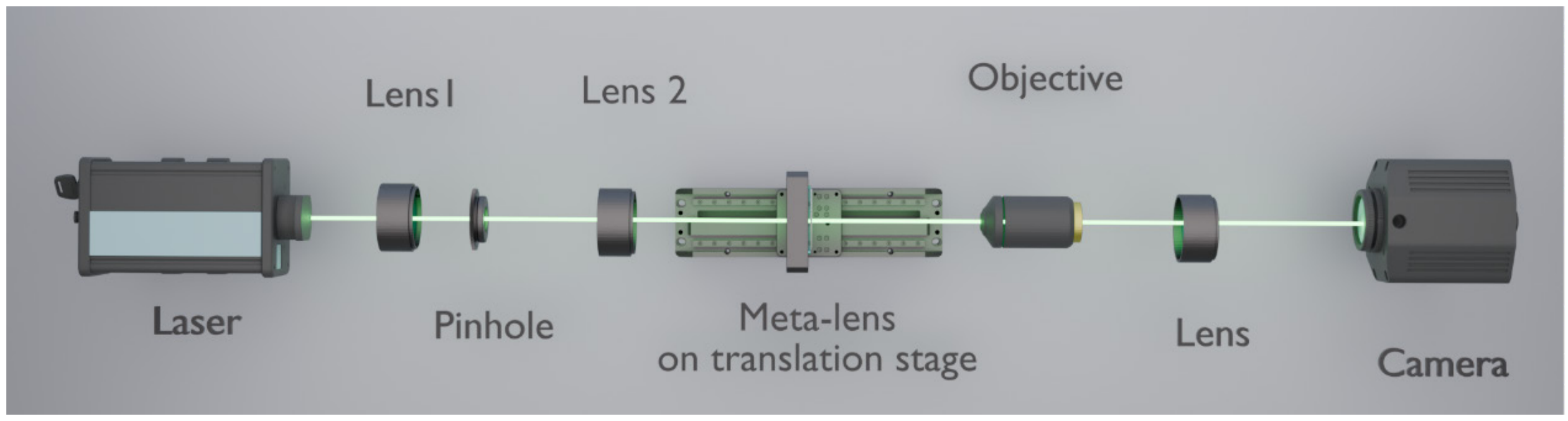

**Figure S2. The experimental setup for characterization of $Si_3N_4$ metalens.**

For USAF 1951 resolution testing, the setup was configured as in Fig. S3. The supercontinuum source was replaced by a quartz–tungsten–halogen lamp (Thorlabs QTH10/M). The illumination was first condensed and then collimated (L1/Obj.1 to concentrate the light, L2 to produce a parallel beam). A band-pass filter (550 ± 30 nm) set the test wavelength. A diffuser was inserted to homogenize the intensity and angular distribution across the field. The collimated beam illuminated the resolution target, placed at the focal plane of the metalens. The metalens-formed image was collected by a 10× objective (NA = 0.26) and relayed by a 200 mm lens; with the flip mount mirror M1 removed, the image propagated straight to CCD camera 2, where the resolution patterns were recorded.

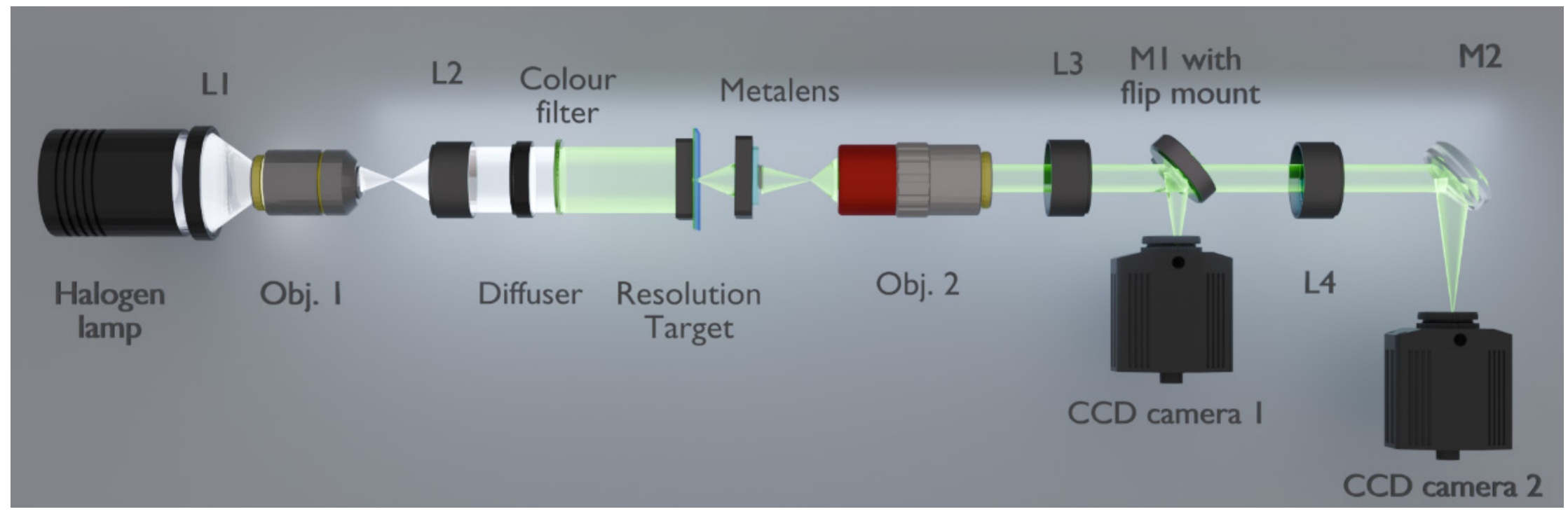

**Figure S3. The experimental setup for measurement for the 1951 USAF resolution test chart and capture images with $Si_3N_4$ metalens.** With M1 removed, the beam propagates directly to CCD camera 2 for resolution-chart measurements; with M1 inserted, the beam is redirected to the object-imaging arm and recorded on CCD camera 1**.**

For real-object imaging, M1 was flipped in. The resolution target was replaced with the object (image slide). The rest of the optics were kept the same. With M1 inserted, the beam was directed into the measurement arm used for recording the object image on CCD camera 1.

## Supplementary Note 3: Representative simulated imaging patterns for USAF resolution targets

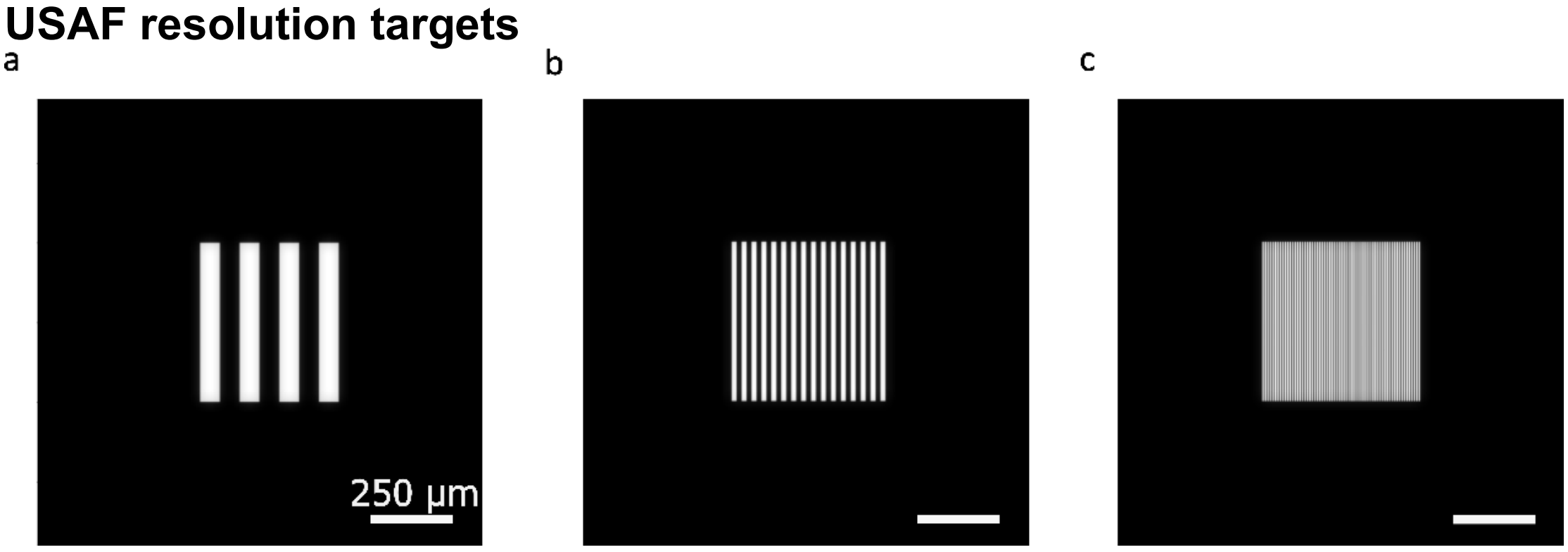

**Figure S4 Representative simulated images of USAF resolution targets are shown for Group 3 Element 1, Group 5 Element 1, and Group 7 Element 1 respectively (from left to right).** All images were obtained using the same imaging model and system parameters under incoherent illumination at a single wavelength. The images have the same scale bar.

The USAF-1951 imaging patterns were simulated using a scalar Fourier-optics model to provide a diffraction-based reference for comparison with the experimentally measured USAF images. The model included the finite aperture of the metalens, the object–image geometry used in the experiment and CCD pixel sampling, but did not include fabrication-induced phase errors, local scattering, residual alignment errors or experimental noise.

The object was modelled as a USAF-1951 resolution target with spatial frequencies defined by the corresponding group and element indices. For a USAF element with group number $G$ and element number $E$, the spatial frequency is

$$\nu_{G,E}=2^{G+\frac{E-1}{6}}\,\mathrm{lp\ mm^{-1}}, \tag{1}$$

Each USAF element was represented by a finite three-line-pair binary intensity pattern with the corresponding spatial period,

$$p_{G,E}=\frac{1}{\nu_{G,E}} \tag{2}$$

The metalens was modelled as an ideal phase-only transmission function designed to focus normally incident light to a focal length of $f = 4\ mm$ at $\lambda = 550nm$. The ideal focusing phase profile is given by

$$\phi(r)=-\frac{2\pi}{\lambda}\left(\sqrt{r^2+f^2}-f\right) \tag{3}$$

where $r=\sqrt{x^2+y^2}$ is the radial coordinate. The finite aperture was included using a circular pupil function,

$$P_{(r)}=\begin{cases}1, r\leq R,\\ 0, r>R,\end{cases} \tag{4}$$

where $R$ is the effective aperture radius. The complex pupil function of the ideal metalens is therefore

$$t_{(r)}=P_{(r)}\exp[i\phi(r)]. \tag{5}$$

The numerical aperture is given by

$$\mathrm{NA}=\frac{R}{\sqrt{f^2+R^2}} \tag{6}$$

which defines the spatial-frequency support of the imaging system. For incoherent imaging, the diffraction-limited cutoff spatial frequency is

$$\nu_c = \frac{2NA}{\lambda} \tag{7}$$

Free-space propagation between planes was calculated using the angular-spectrum method. For an input complex field $U(x, y; 0)$, the propagated field at distance $z$ is

$$\mathrm{U(x, y; z)} = \mathcal{F}^{-1}\left[\mathcal{F}\{\mathrm{U(x, y; 0)}\}\exp\left(\mathrm{iz}\sqrt{\mathrm{k^2 - k_x^2 - k_y^2}}\right)\right] \tag{8}$$

where $\mathcal{F}$ and $\mathcal{F}^{-1}$ denote the two-dimensional Fourier transform and inverse Fourier transform, respectively, $k = 2\pi / \lambda$ , and $k_x$ and $k_y$ are transverse spatial-frequency coordinates. Evanescent components outside the propagating spatial-frequency region were excluded.

Under incoherent illumination, image formation was described as a convolution between the object intensity and the incoherent intensity point-spread function,

$$\mathrm{I(x, y) = O(x, y) * PSF(x, y)} \tag{9}$$

where $O(x, y)$ is the object intensity and $PSF(x, y)$ is the intensity point-spread function of the system. The PSF was calculated from the coherent impulse response and normalized to conserve energy.

$$\iint \mathrm{PSF(x, y)\ dx\, dy} = 1 \tag{10}$$

For comparison with the experiment, the imaging PSF was calculated using the same object distance and image-plane distance as those used in the measurement. The simulated image therefore includes the magnification-dependent blurring associated with the selected object–image geometry, rather than an ideal focal-plane PSF alone. After convolution, the simulated USAF image was sampled using the CCD pixel sampling used in the experiment. Quantitative CTF extraction was performed on the sampled image. Display normalization was applied only for visualization and was not used in the CTF calculation.

The contrast transfer function was quantified from the USAF images by extracting the modulation depth of the line-pair patterns. For each spatial frequency, five rectangular regions of interest were selected within the corresponding USAF element. The ROIs were placed at different positions along the bar direction while avoiding the edges of the patterned region, transition regions between adjacent elements and obvious local defects or background artefacts. The same ROI-selection strategy was applied to the experimental and simulated images.

For each ROI, the intensity was averaged along the direction parallel to the bars to obtain a one-dimensional intensity profile perpendicular to the line pairs. For vertical bar patterns, this averaging is written as

$$\bar{\mathrm{I}}_{\mathrm{i}}(\mathrm{x}) = \frac{1}{\mathrm{N}_{\mathrm{y\in ROI_i}}}\sum\nolimits_{\mathrm{y\in ROI_i}} \mathrm{I\,(x, y)} \tag{11}$$

where $i$ denotes the ROI index and $N_y$ is the number of pixels along the averaging direction. This averaging reduces pixel-level noise while preserving the modulation across the bright and dark bars.

The CTF of the $i$-th ROI was calculated using the Michelson contrast,

$$\mathrm{CTF_i} = \frac{\mathrm{I_{max,i} - I_{min,i}}}{\mathrm{I_{max,i} + I_{min,i}}}, \tag{12}$$

where $\mathrm{I_{max,i}}$ and $\mathrm{I_{min,i}}$ are the bright- and dark-line intensities extracted from the averaged one-dimensional profile. For low-spatial-frequency elements, $I_{max,i}$ and $I_{min,i}$ were obtained by averaging the bright and dark plateau regions. For high-spatial-frequency elements, where the profile approached a sinusoidal modulation, local maxima and minima were used. The peak-detection distance was adjusted according to the expected line-pair period in pixels to avoid merging neighbouring periods or detecting noise-induced peaks.

For each spatial frequency, the reported CTF was calculated as the mean value across five ROIs. The error bar was calculated as the sample standard deviation,

$$\sigma_{\mathrm{CTF}} = \sqrt{\frac{1}{4}\sum_{i=1}^{5}\left(\mathrm{CTF_i} - \overline{\mathrm{CTF}}\right)^2} \quad (13)$$

Thus, the error bars in Fig. 2e represent $\overline{\mathrm{CTF}} \pm \sigma_{\mathrm{CTF}}$, not the standard error of the mean. The simulated CTF was extracted using the same procedure. Because the numerical model was spatially uniform and contained no experimental noise, the ROI-to-ROI variation in the simulated CTF was negligible; simulated error bars are therefore omitted from Fig. 2e for clarity.

## Supplementary Note 4: Extended phantom validation across varying preload strains

Figures S5a and S5b present the extended results for the 2 mm cylindrical inclusion under sequential axial preloads of 10% and 20%, respectively. At a minimal preload of 10% (Fig. S5a), the raw phosphorescent emission modulation [panel (ii)] exhibits localised but subtle intensity gradients. Despite this low-input deformation regime, the MSOP system successfully resolves the embedded inclusion, yielding a high-fidelity reconstructed stress map [panel (iii)] that clearly differentiates the stiff inclusion from the compliant background. As the axial preload increments to 20% (Fig. S5b), the signal-to-noise ratio of the raw sensing layer improves substantially due to increased contact pressure. This amplification of the input mechanical signal translates directly into the reconstructed stress maps [panel (iii)], manifesting as an elevated stress amplitude and a progressive sharpening of the boundary gradients. The corresponding horizontal and vertical stress profiles [panels (iv) and (v)] map this transition quantitatively. The first-order Gaussian derivative fits (red dashed lines) demonstrate tight convergence to the true physical boundaries (yellow dashed lines) across both loading regimes, confirming that the edge-localisation capability remains highly accurate even as non-linear strains scale.

The accompanying violin plots [panel (vi)] confirm this strain-dependent behaviour; as the macroscale deformation increases from 10% to 20%, the continuous pixel-intensity distribution within the inclusion domain shifts upward and separates distinctively from the surrounding compliant matrix baseline, tracking the predictable mechanical response of the silicone structures. Furthermore, across both loading states, the edge-localisation derivative profiles reveal highly symmetrical lateral expansions along both the horizontal and vertical axes during compression. This isotropic Poisson expansion contrast indicates excellent spatial alignment and confirms that the automated edge-detection framework remains stable and immune to minor experimental loading asymmetries or micro-scale tilts during physical compression. Ultimately, these results validate that the MSOP platform delivers a highly robust signal-to-noise ratio and exceptional boundary-tracking fidelity across varying operational strain regimes.

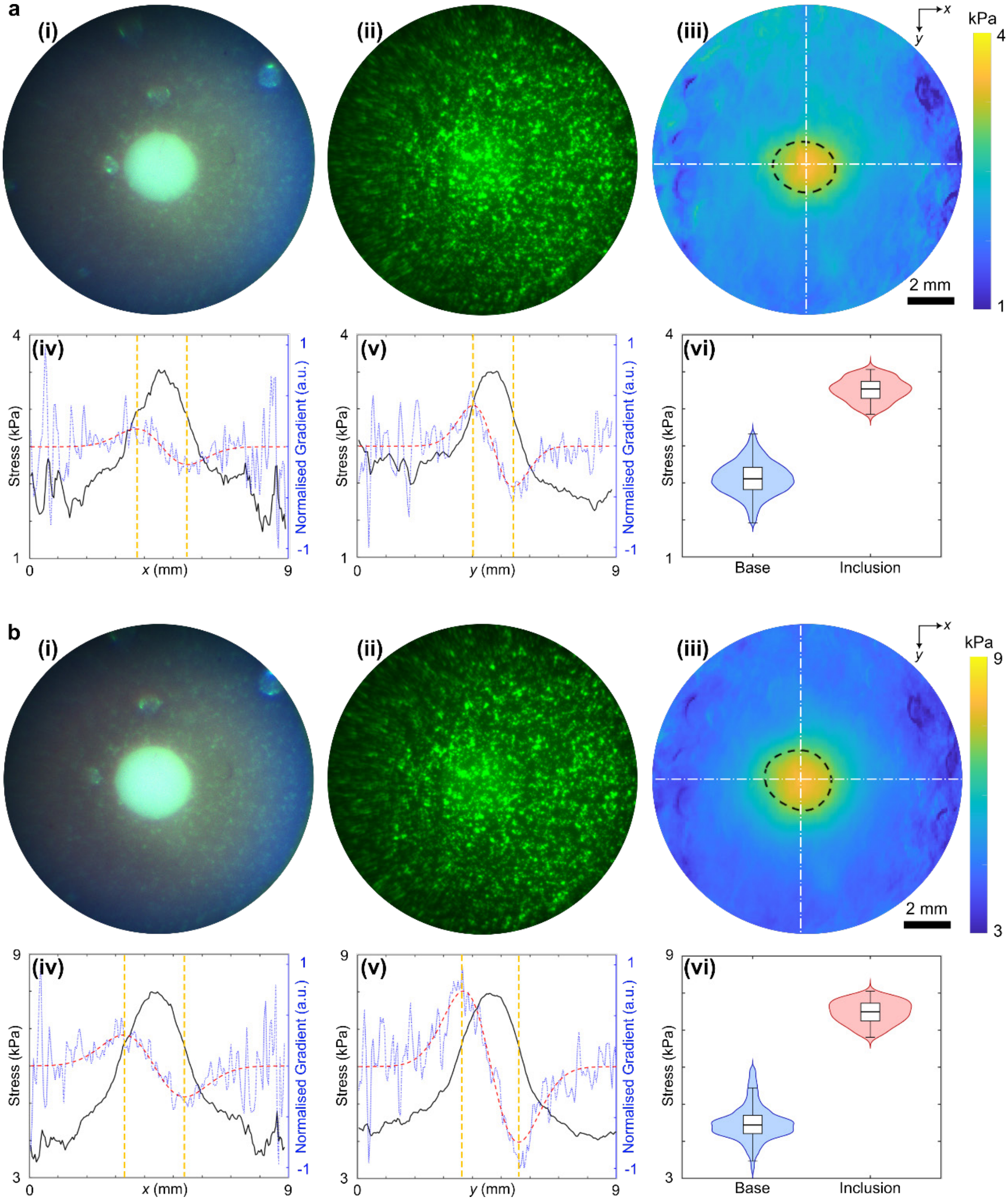

**Figure S5**. **Extended phantom validation.** Performance of the MSOP system on a silicone phantom with a 2 mm diameter stiff inclusion under 10% (**a**) and 20% (**b**) axial preload strain. For each set, sub-panels (i–iii) show the white-light morphology, filtered UV-light images of the stress sensing layer, and the reconstructed stress maps, respectively. Sub-panels (iv, v) present transverse stress profiles (white lines) along the horizontal and vertical axes [marked in (iii)], with corresponding stress derivatives (blue dashed lines) and first-order Gaussian derivative fits (red dashed lines) used for edge localisation (yellow dashed lines). Sub-panel (vi) shows the statistical distribution of measured stress within the inclusion [marked by dashed frame in (iii)] and base matrix, visualised via violin plots. In the violin plots, the box-and-whisker elements represent the median, interquartile range, and range (minimum to maximum).

## Supplementary Note 5: Extended results of the mouse pancreatic cancer specimens

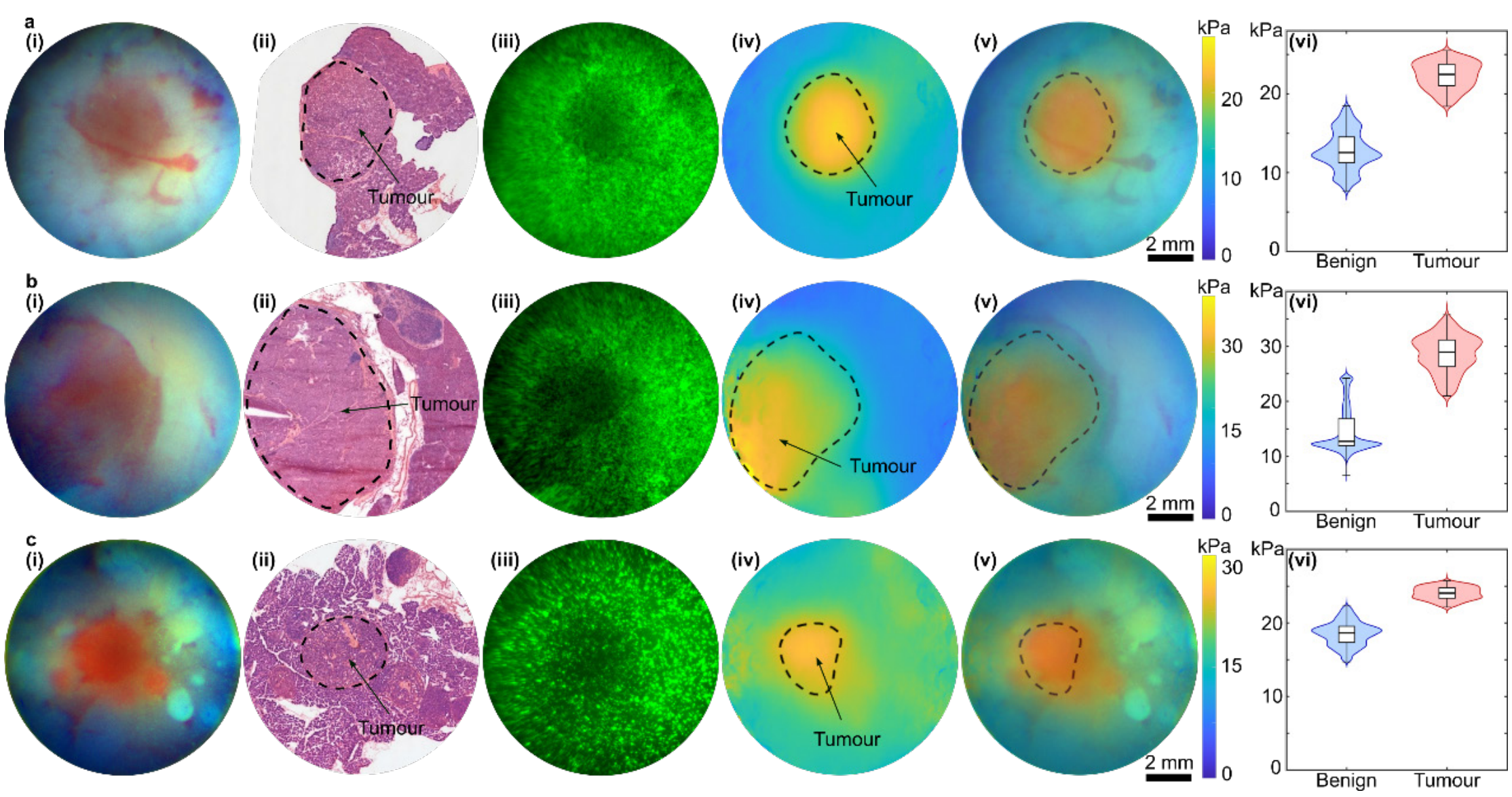

**Figure S6**. **Extended mouse pancreatic cancer results. a-c** Multi-modal imaging and mechanical characterisation of three mouse pancreatic specimens. Sub-panels (i–vi) show the MSOP white-light morphology, H&E-stained histology (black dashed frame indicates tumour region), filtered UV-light images, reconstructed stress maps, stress/white-light overlays (black dashed curve indicates detected stress boundary), and stress distributions (tumour in red; benign in blue).